\documentclass[10pt]{article}
\usepackage[sectionbib]{natbib}
\usepackage{array,epsfig,fancyheadings,rotating}
\usepackage{hyperref}

\usepackage{amsmath}
\usepackage{amssymb}
\usepackage{amsfonts}
\usepackage{multirow}
\usepackage{amsthm}
\usepackage{times}
\usepackage{authblk}
\usepackage{booktabs}
\usepackage{graphicx}
\usepackage{epsfig,color}

\theoremstyle{definition}

\newcommand{\inner}[2]{\left\langle #1,#2 \right\rangle}

\newcommand{\real}{\ensuremath{\mathbb{R}}}
\newcommand{\s}{\ensuremath{\mathbb{S}}}
\newcommand{\ltwo}{\ensuremath{\mathbb{L}^2}}

\newcommand{\D}{{\mathcal D}}
\newcommand{\bm}[1]{\boldsymbol{#1}}
\DeclareMathOperator{\bc}{\boldsymbol{c}}
\newcommand{\ra}[1]{\renewcommand{\arraystretch}{#1}}
\usepackage{authblk}

\title{Robust Comparison of Kernel Densities on Spherical Domains}
\author[1]{Zhengwu Zhang\thanks{zhengwu\_zhang@urmc.rochester.edu}}
\author[2]{Eric Klassen\thanks{klassen@math.fsu.edu}}
\author[3]{Anuj Srivastava\thanks{anuj@stat.fsu.edu}}
\affil[1]{Department of Biostatistics and Computational Biology, University of Rochester, Rochester, NY}
\affil[2]{Department of Statistics,  Florida State University, Tallahassee, FL}
\affil[3]{Department of Mathematics,  Florida State University, Tallahassee, FL}

\begin{document}

\maketitle




\renewcommand{\baselinestretch}{1.2}

\markright{ \hbox{\footnotesize\rm 
}\hfill\\[-13pt]
\hbox{\footnotesize\rm
}\hfill 
}

\markboth{\hfill{\footnotesize\rm Z. ZHANG, E. KLASSEN, AND A. SRIVASTAVA} \hfill}
{\hfill {\footnotesize\rm DENSITY COMPARISON AND TWO-SAMPLE TEST} \hfill}

\renewcommand{\thefootnote}{}
$\ $\par




\begin{quotation}
\noindent {\it Abstract:}

While spherical data arises in many contexts, 
including in directional statistics, the current tools for density
estimation and population comparison on spheres are quite limited. 
Popular approaches for comparing populations (on Euclidean domains) mostly involve
a two-step procedure: (1) estimate {\it probability density functions} ({\it pdf}s) from their respective samples, 
most commonly using the kernel density estimator, and, (2) compare {\it pdf}s using a metric such as the $\ltwo$ norm. 
However, both the estimated {\it pdf}s and their differences depend heavily on the chosen kernels, 
bandwidths, and sample sizes. 
Here we develop a framework for comparing spherical populations that is robust to these choices. 
Essentially, we characterize {\it pdf}s on spherical domains by quantifying their smoothness. Our framework uses a spectral representation, 
with densities represented by their coefficients with respect to the eigenfunctions of the Laplacian operator on a sphere. The change in smoothness, 
akin to using different kernel bandwidths, is controlled by exponential decays in coefficient values. 
Then we derive a proper distance for comparing
{\it pdf} coefficients while equalizing smoothness levels, negating influences of sample size and bandwidth.  
This signifies a fair and meaningful comparisons of populations, despite vastly different sample sizes,
and leads to a robust and improved performance.  
We demonstrate this framework using examples of variables on $\s^1$ and $\s^2$,  and evaluate its performance 
using a number of simulations and real data experiments.\par

\vspace{9pt}
\par
\end{quotation}\par

\section{Introduction}

\begin{figure}
\begin{center}
\footnotesize
\begin{tabular}{|cc|}
\hline
\includegraphics[height=1.0in]{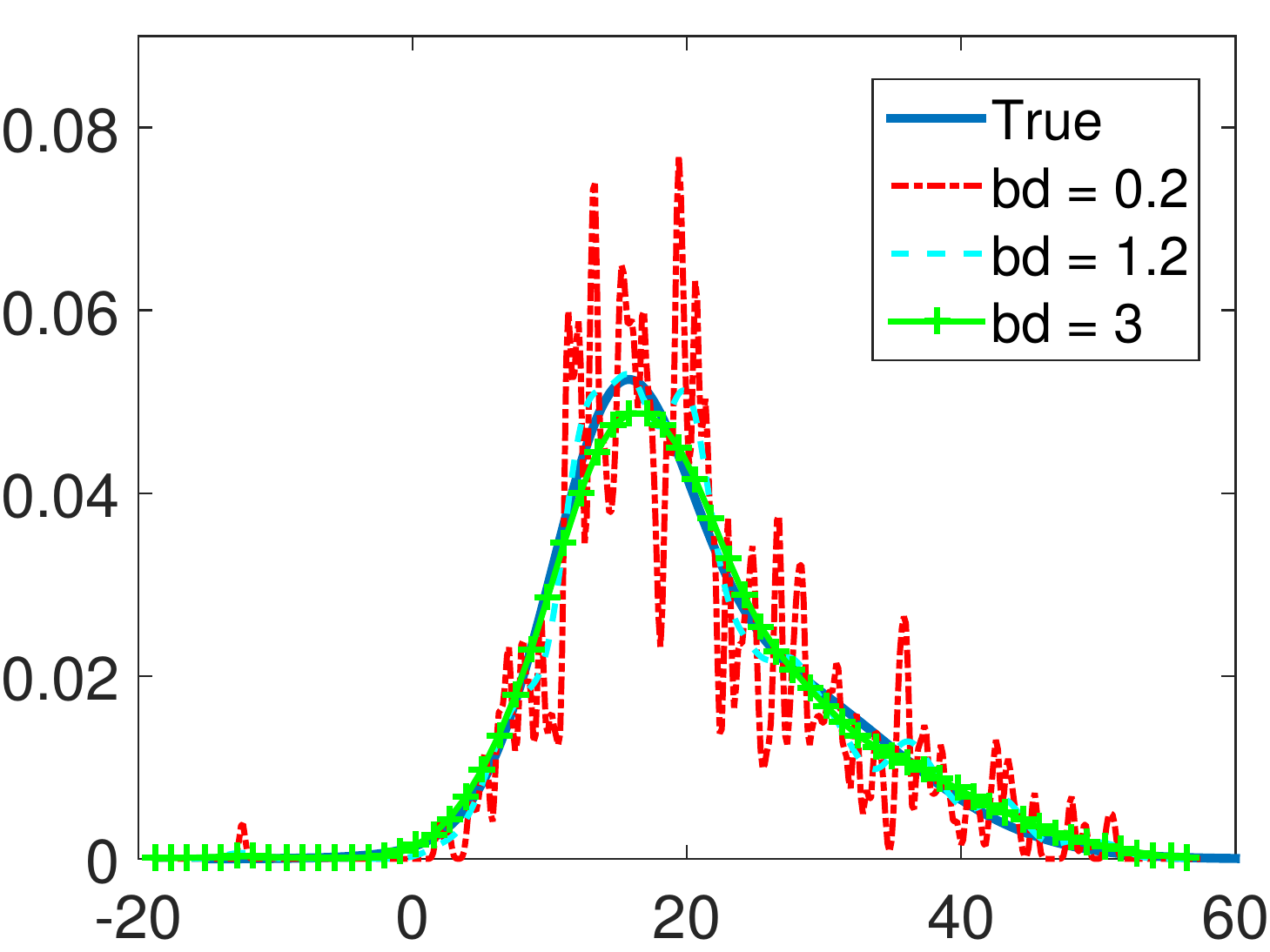}&
\includegraphics[height=1.0in]{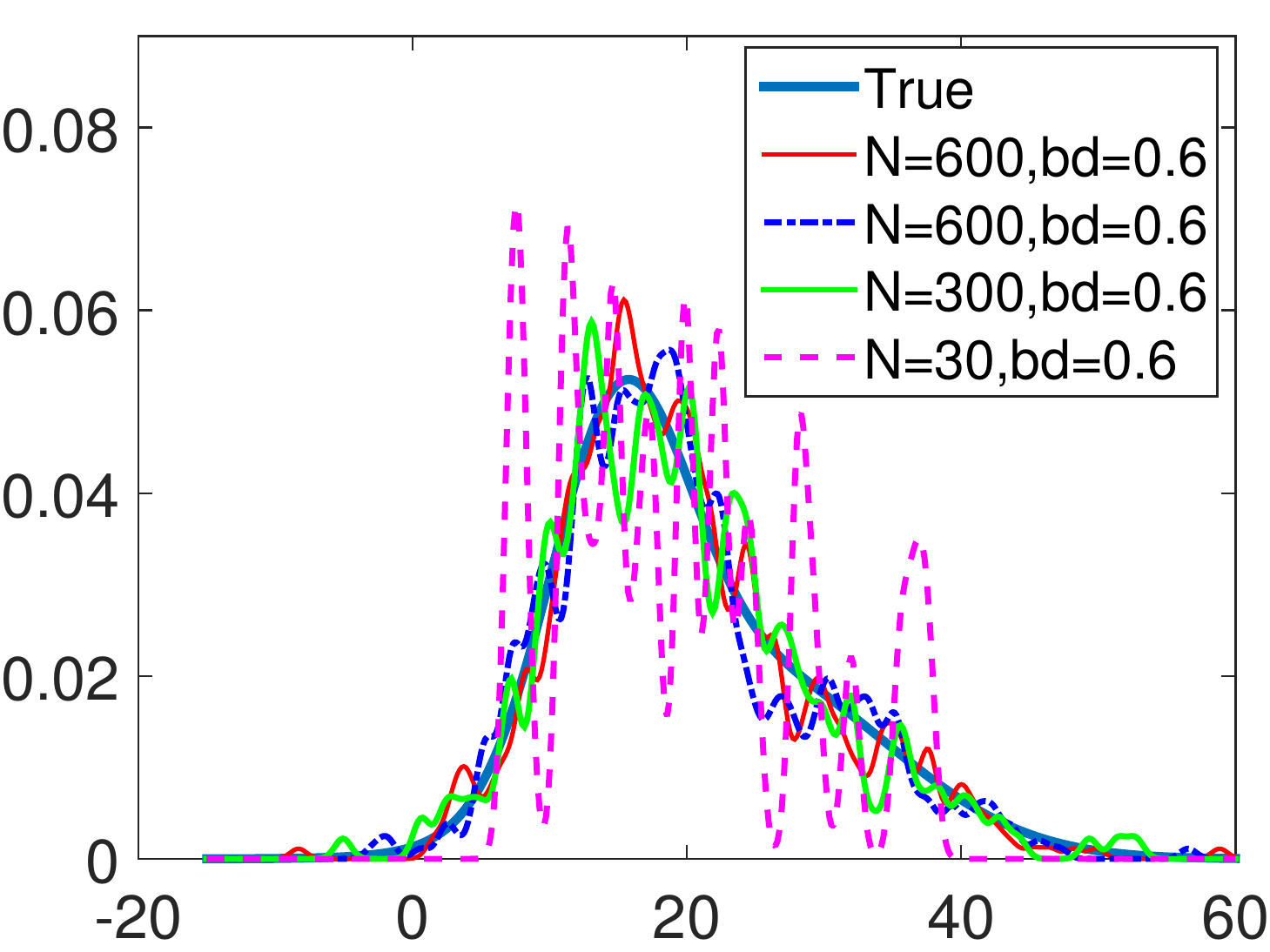}\\
(a) & (b) \\
\hline
\includegraphics[height=1.0in]{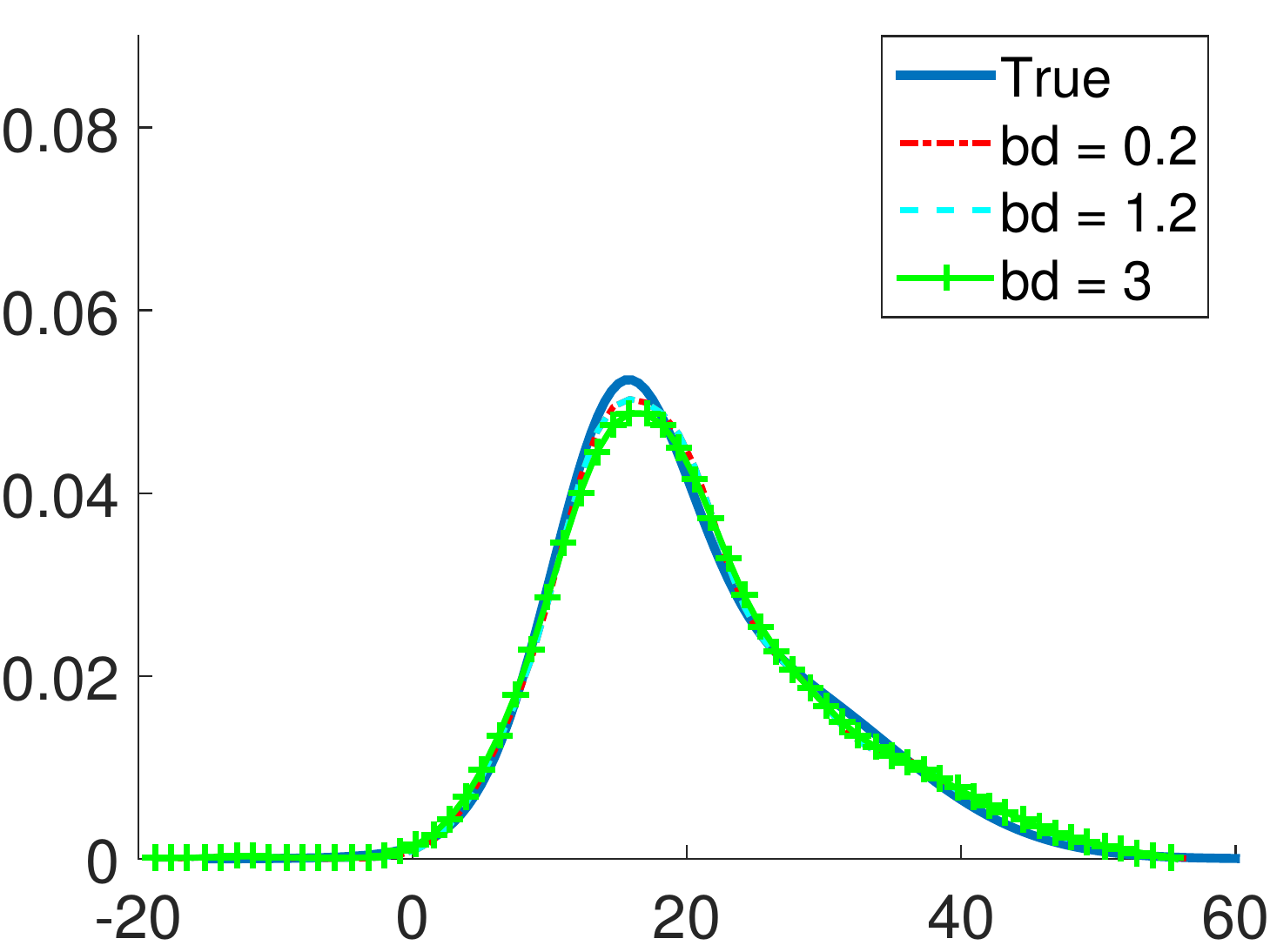}&
\includegraphics[height=1.0in]{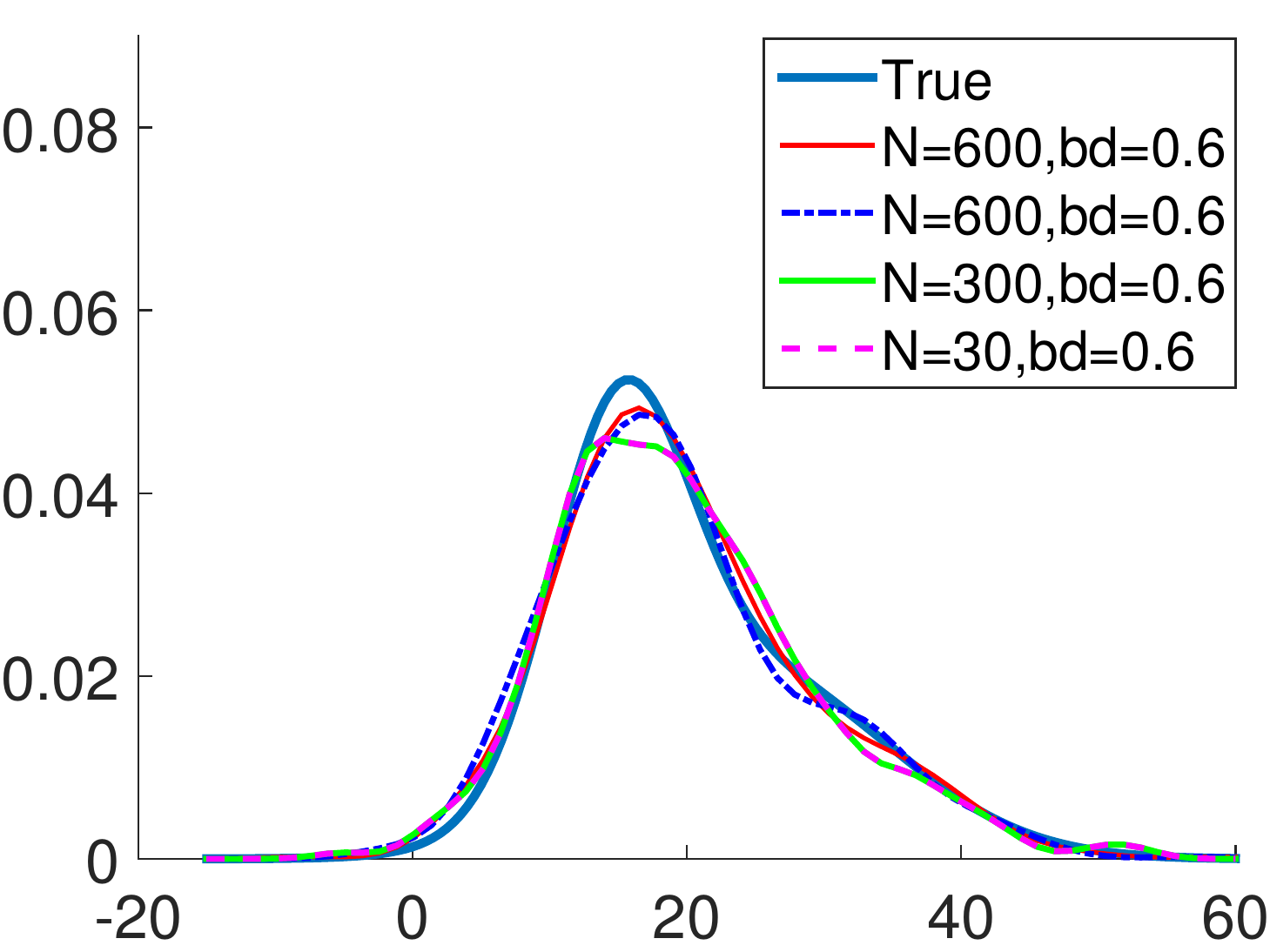}\\
(c) & (d) \\
\hline
\end{tabular}
\caption{Examples of kernel density estimation. (a) and (b) Kernel estimates under different bandwidths and sample sizes. (c) and (d) Estimated densities in (a) and (b) at the same smoothness level as the true density.  Here, ``bd'' indicates the bandwidth and ``N'' indicates the sample size.
}
\label{fig:bandwidthsize}
\end{center}
\end{figure}

The estimation of probability density functions ({\it pdf}s) and comparisons of underlying populations 
are fundamental problems in statistics. In a variety of 
situations, where data satisfy some natural constraints,  it is better to view and analyze data as elements of a non-Euclidean manifold. 
A simple example is directional 
statistics, where one deals with analysis of data on a unit sphere. In order to understand the limitations of current 
solutions, for estimating and comparing densities on spherical domains, we start with a discussion of methods in Euclidean domains. 
The classical nonparametric estimate of a {\it pdf}, given samples from that density, 
is a kernel density estimate \citep{RosenblattM1956,Parzen:1962}. This approach is commonly used for Euclidean domains but
can be easily adapted to spheres also. There are two key choices to be made in this 
estimation: (1) the kernel function, a symmetric unimodal function that integrates to one, and (2) the bandwidth.  
It is widely acknowledged that the choice of bandwidth is more influential than the choice of kernel in terms of {\it pdf} estimation performance. 
Henceforth, in this paper, we will fix the kernel to be an isotropic (i.e., circularly symmetric) Gaussian-type kernel and focus 
on the issues arising from using different bandwidths. The choice of Gaussian kernel facilitates a group structure that will be exploited later in this paper.
To highlight the importance of bandwidth in density estimation, 
Figure \ref{fig:bandwidthsize} shows an example of {\it pdf} estimation in $\real^1$. The 
panel (a) shows several estimates of the {\it pdf} for different bandwidths on the same data. Another factor that drastically affects the final estimate is the sample size, 
as highlighted in Figure  \ref{fig:bandwidthsize} panel (b).

Our interest in this paper is more on comparing populations rather than just estimating {\it pdf}. If we use kernel density estimates 
and compare them using one of standard metrics,  the 
results will naturally be very sensitive to the choice of bandwidths and sample sizes. In order to make this comparison robust
to low sample size and different bandwidth choices, there are several possibilities:  
\begin{enumerate} 
\item {\bf Use a fixed bandwidth.} We can fix a bandwidth for all {\it pdf} estimates, and then use any function norm \citep{cha2007} for
comparison.  While this is a convenient strategy, it suffers from the problem that 
the final answer will strongly depend on the sample size \citep{ECT:7906343} (also illustrated in Figure \ref{fig:bandwidthsize}). 
Different sample sizes can lead to very different {\it pdf} estimates when using the same bandwidth despite coming from the same underlying 
distribution.  

\item {\bf Use an adaptive bandwidth.} We can use one of 
bandwidth selection methods \citep{Jone1996,Jone19962,Bowman1984,scott1987biased,turlach1993,Botev2010} 
to estimate {\it pdf}s and then compare them. However, there is no consensus on 
which approach works best in general scenarios. Most bandwidth selection methods are based on 
minimizing the integrated squared error or the mean integrated squared error (MISE), but 
they often fail in practice because the true {\it pdf}  that is necessary for calculating these quantities  is unknown. 
\item {\bf Use a fixed smoothness level.} Another solution, coming from a very different perspective, is to focus on the smoothness of the 
estimated {\it pdf}s rather than on the bandwidth, which is the main idea of this paper. We want to quantify the level of smoothness
of a {\it pdf} as a function and use that in the following way. For any two estimated {\it pdf}s being 
compared,  one can bring them to the same level of smoothness, irrespective of their initial bandwidths and sample sizes. 
Since even the classical estimation theory makes assumption about smoothness of underlying density \citep{Marron1988,chaudhuri2000}, it is a natural criterion to 
include in estimation. 
Furthermore, this property can be easily manipulated, as described later, and provides robustness against the choice of bandwidths and 
sample sizes. 
Figure  \ref{fig:bandwidthsize} bottom row shows {\it pdf} estimates from the top row after they have been processed to
equalize their smoothness level
(details of this process are presented later). Now these estimates appear far more similar to each other than before, as they should be.
\end{enumerate}

While comparisons of populations are needed everywhere, we consider two important applications. The first application is in computer vision and image analysis, where a variety of image analysis techniques rely on specifying certain features of interest, e.g., Haar \citep{Viola2001}, HOG \citep{DT05}, SIFT \citep{David} and so on, and comparing differences in estimated densities of these features. The second application is the two-sample hypothesis testing. 
Any measure of difference between estimated densities is a natural statistic for two-sample test,  e.g. Kolmogorov-Smirnov (KS) test \citep{Smirnov1948}. 
Such methods depend
on kernel density estimates,  and the 
bandwidth parameter strongly influences final results.   It will be very useful to develop a metric that depends on 
something that is more intrinsically related to the underlying densities rather than the bandwidth parameter, and is robust to 
variability in sample size.

The kernel-based density estimation is essentially a problem of smoothing data. Given a random sample $\{x_1,x_2,...,x_T\}$, the empirical density function is given by $f^E(x) = T^{-1}\sum_{i=1}^T \delta(x-x_i)$, where $\delta(x)$ denotes a point mass at $x$. The convolution of $f^E(x)$ with a kernel function $K_h(x)$ gives us 
an estimated {\it pdf} $(f^E * K_h)(x) = T^{-1} \sum_{i=1}^T K_h(x-x_i)$, where $*$ denotes the convolution operation. If $K_h(x)$ is the Gaussian kernel with a bandwidth $h$, the convolution process is called {\it Gaussian smoothing} or {\it blurring}. This smoothing is similar to the Gaussian blur of images \citep{zhang2013}. 
As described there, one can study Gaussian blur as a solution of heat diffusion equation with appropriate initial condition. It turns out that the set of all 
isotropic Gaussian kernels, under all possible bandwidths, form a {\it group}. The {\it orbit} of density functions under this group action 
defines an equivalence class; in the current context, it can be viewed as the set of all {\it pdf}s estimated from the same data but 
with different bandwidths. This solution naturally applies to spherical domains also and is therefore 
a good solution for density estimation and population comparison on a sphere. 
 
The novel contributions of this paper are as follows. (1) Given
kernel density estimates on a spherical domain, estimated using Gaussian kernel with arbitrary bandwidths, 
our framework identifies the equivalence classes to which they belong. 
It then compares these estimates by comparing their equivalence classes, and thus is robust to the 
original bandwidth parameter.  (2) We define a function $G$ that  quantifies smoothness of {\it pdf}s, and 
use it to specify the {\it section} of action of the blurring group. 
Two functions are in the same section if they have the same level of smoothness. 
(3) This framework is applied to develop a two-sample hypothesis test where {\it pdf}s estimated from data with 
arbitrary sample sizes are brought to the 
same smoothness level, i.e., the same section, and then compared via a manifold distance. 

The rest of the paper is organized as follows. In Section 2, we lay out the mathematical foundation of our approach. In Section 3, we 
apply this framework to the kernel density estimation and population comparison on different spherical domains. 
In Section 4, we develop a two-sample hypothesis test and in Section 5
we provide a variety of experimental results using simulated and real data.

\section{Mathematical Framework}
We start by outlining mathematical details of our framework, including 
Gaussian heat kernel, kernel density estimation, bandwidth selection, and a metric for comparing estimated densities. 

\subsection{Heat Equation for Density Estimation}

Let $\mathcal{F}$ denote the set of smooth, non-negative functions on a domain $\D$, 
and $\mathcal{F}_0$ be the subset of {\it pdf}s. That is,  $\mathcal{F} = \{f: \D \rightarrow \real^+ |f \text{ is smooth}\}$, and
$\mathcal{F}_0 = \{f \in \mathcal{F} |  \int_\D f = 1 \}$.
Let $L: \mathcal{F} \rightarrow \mathcal{F}$ be the standard Laplacian operator on $\mathcal{F}$.  
In this paper, we consider the compact domains such as  $\D = \s^1$ (a circle), $\s^1 \times \s^1$ (a torus)  and $\s^2$ (a sphere).   
For each of these compact domains, it is easy to find an orthonormal Hilbert basis of $\ltwo(\D,\real)$ with the property that every 
basis element is an eigenfunction of the Laplace operator. 
Extension to Euclidean domains such as $\real^1$ and $\real^2$ will be also discussed later 
although the Laplacian operator is defined differently in such non-compact domains. 
For $\D = \s^1$, we have $L\cdot f = - \frac{\partial^2 f}{\partial x^2}$; for $\D = \s^1 \times \s^1$,  $L\cdot f = - \frac{\partial^2 f}{\partial x_1^2} -\frac{\partial^2 f}{\partial x_2^2}$ for ${\bm x}=(x_1,x_2)$, and for $\D = \s^2$,  $L \cdot f = -\frac{1}{\sin^2 \varphi}\frac{\partial^2}{\partial \theta^2} - \frac{1}{\sin \varphi}\frac{\partial}{\partial \varphi}(\sin \varphi \frac{\partial}{\partial \varphi})$ for the spherical coordinate ${\bm x}=(\theta,\phi)$, where $\theta \in [0,\pi]$ is the polar angle and $\varphi \in [0, 2\pi)$ is the azimuthal angle.

The kernel density estimator based on sample data $\{x_1,x_2, \cdots, x_T\}$ with $x_i \in \D$ is 
$\hat{f}_h(x) = T^{-1} \sum_{i=1}^T K_h(x-x_i)$,
where $K_h(x)$ is the Gaussian kernel on $\D$,  and $h \in \real^+$ is the bandwidth. If we treat the bandwidth as time, 
the smoothness of the estimated density will increase as the time increases. Another way to state this is to use the classical heat diffusion equation:
\begin{equation} \label{eqn:heateq}
\frac{\partial f(t,x)}{\partial t} = - (L\cdot f)(t,x),
\end{equation}
where $L$ is the Laplacian operator.  {The Gaussian kernel $K_h(x)$ used in this paper needs to satisfy this heat equation \citep{hartman1974} }.  
In Eqn. (\ref{eqn:heateq}), the value of $t$ has the same effect as $h$ in the kernel estimate, 
and therefore, the time parameter in heat diffusion resembles the bandwidth in kernel density estimation. 
If we set the initial heat to be a given function say $f_0(\mathbf{x}) \in \mathcal{F}_0$ (say
a kernel density estimate using a bandwidth $h_0$), the solution $f(t,\cdot)$ for $t>0$ is also a kernel density estimate with a larger bandwidth $h_0 + h$ for some $h>0$. For more mathematical details on the heat equation, readers are referred to \cite{Lindeberg90} and \cite{chaudhuri2000}.

We represent a smooth {\it pdf} $f_0 \in \mathcal{F}_0$ on the domain $\D$ via its coefficients under a complete orthonormal basis set. Assuming that the domain $\D$ is a compact domain, e.g., $\D=\s^1$, and using the $\ltwo$ metric on $\mathcal{F}$, we
define a complete orthonormal Hilbert basis $\{\phi_0,\phi_1,\phi_2,...\}$, where each $\phi_n$ is an eigenfunction of $L$ with eigenvalue $\lambda_n$, i.e., $L \cdot \phi_n = \lambda_n \phi_n$. Assuming that $\phi_0$ is a constant function, 
we have $\lambda_0=0$, and all other $\lambda_n$s are positive due to the positive definiteness of $L$.
Any element $f_0 \in \mathcal{F}_0$ can then be expressed as $f_0(x) = \sum_{n=0}^\infty c_n \phi_n(x)$. 
In practice, we use a basis set of size $N < \infty$ to make this representation finite. 
So $f_0$ is (approximately) represented by a vector $\bc \equiv \{c_n, n = 0,...,N \} \in \real^{N+1}$. Note that for a rough density function, one may need a large $N$ to 
more accurately represent the function.  {We define a mapping $\Pi: \mathcal{F} \mapsto \real^{N + 1}$, i.e., $\Pi(f) = \bc$ for $\bc \in \real^{N + 1}$. 
As specified thus far, $\Pi$ is a many-to-one map meaning that its inverse is set-valued. 
However, we will use $\Pi^{-1}(\bc)$ to denote a specific density function given by $\sum_{n=0}^N c_n \phi_n(x)$ (because of the 
constraint of a density function, we slightly adjust $c_0$  such that $\int_D \sum_{n=0}^N c_n \phi_n(x) dx = 1 $).}

The advantage of the chosen basis is that after expressing functions with coefficients under this basis, 
we can easily express the solution for the heat equation analytically.  If $f(t,x)$ is the solution of the heat equation, with the initial heat distribution $f_0(x)=\sum c_n \phi_n(x)$, 
then this solution takes the form $f(t,x) = \sum_{n=0}^N e^{-\lambda_n t}c_n \phi_n(x)$. Using simple calculus one can verify that the right part of the heat equation is 
$$- L\cdot f(t,x) = \sum_{n=1}^N e^{-\lambda_n t}c_n (L \cdot \phi_n) = -\sum_{n=1}^N e^{-\lambda_n t} c_n \lambda_n \phi_n,$$ 
and the left part of the heat equation is
$$ \frac{\partial f(t,x)}{\partial t} = - \sum_{n=0}^N \lambda_n e^{-\lambda_n t}c_n \phi_n\ .
$$ 
Therefore, $-(L \cdot f)(t,x)$ exactly equals to ${\partial f(t,x)}/{\partial t}$, and $f(t,x) = \sum_{n=0}^N e^{-\lambda_n t}c_n \phi_n(x)$ is the solution of the heat equation. In other words, $f(t,x) = \sum_{n=0}^N e^{-\lambda_n t}c_n \phi_n(x) $ is another kernel estimated density with a bandwidth larger than that of $f_0$,
and one can use a vector $\tilde{\bc} \in \real^{N+1}$, where $\tilde{\bc} \equiv  \{\tilde{c}_n, n = 0,...,N| \tilde{c}_n = e^{-\lambda_n t} c_n \}$, to represent the $f(t,x)$.

\subsection{Quantify Smoothness Levels using Sections}
\label{sec:invariantmetric}

Any smooth {\it pdf} can now be (approximately) represented by an element of $\real^{N+1}$. 
We observe that the set $\real$ of smoothing parameter $t$ in Eqn. (\ref{eqn:heateq}) has a natural group structure 
under addition operation (see \citep{Boo03}, Chapter 3), and its action on $\real^{N+1}$ is given by the mapping $\real \times \real^{N+1} \rightarrow \real^{N+1}$: 
\begin{equation} \label{eqn:groupact}
(t,\{c_0, c_1,c_2,...,c_N\}) \rightarrow \{ e^{-\lambda_0 t}c_0, e^{-\lambda_1 t}c_1, e^{-\lambda_2 t}c_2,..., e^{-\lambda_N t}c_N \} \quad.
\end{equation}
For $f_0 \in \mathcal{F}$, and its finite representation $\bc \in \real^{N+1}$, 
the orbit under the group action is:
\begin{equation} \label{eqn:orbit}[\bc] = \{\tilde{\bc} \in \real^{N+1}|\tilde{c}_n = e^{-\lambda_n t}c_n, \forall n, \text{ for some } t\in \real\} \quad. 
\end{equation}
In the kernel density estimation scenario, the group action in Eqn. (\ref{eqn:groupact}) can be understood as follows. We first use a bandwidth $h_0$ to estimate the density (using a Gaussian kernel) and set the estimate as the initial heat, denoted as $f_0$ (represented by a vector  $\{c_n, n = 0,...,N \} \in \real^{N+1}$). For a positive time $t>0$, $f(t,x) \equiv \{e^{-\lambda_0 t}c_0, e^{-\lambda_1 t}c_1,..., e^{-\lambda_N t}c_N\}$ is the kernel estimate with bandwidth $h_0+ |h|$ for some $h$; for a negative time $t<0$, $f(t,x)$ is the kernel estimate with bandwidth $h_0 -|h|$.
The orbit of $f_0$ (defined in Eqn. (\ref{eqn:orbit})) is the set of all possible smoothed versions of $f_0$. It can be deemed as an equivalence class 
for the purpose of comparing densities.  \\


\noindent {\bf Orthogonal Section Under Smoothing Action}: 
Under this geometry, the vector space  $\real^{N+1}$ becomes a disjoint union of orbits (equivalence classes). 
Moving along each orbit, toward the direction of increasing $t$, the kernel estimated densities become smoother
and vice-versa. To compare densities, we compare their orbits, i.e., define a distance between these equivalence classes. 
However, since we do not have any metric under which the group action is by isometries, i.e., the orbits are not 
parallel,  we use the concept of {\it orthogonal section} for comparisons.   
An orthogonal section of  $\real^{N+1}$ under the group action 
is defined to be a set $S$ such that: (1) one and only one element of every orbit $[\bc]$ in $\real^{N+1}$ presents in $S$; (2) the set $S$ is perpendicular to every orbit at the point of intersection. 

\begin{figure}
\begin{center}
\begin{tabular}{cc}
\includegraphics[height=1.5 in]{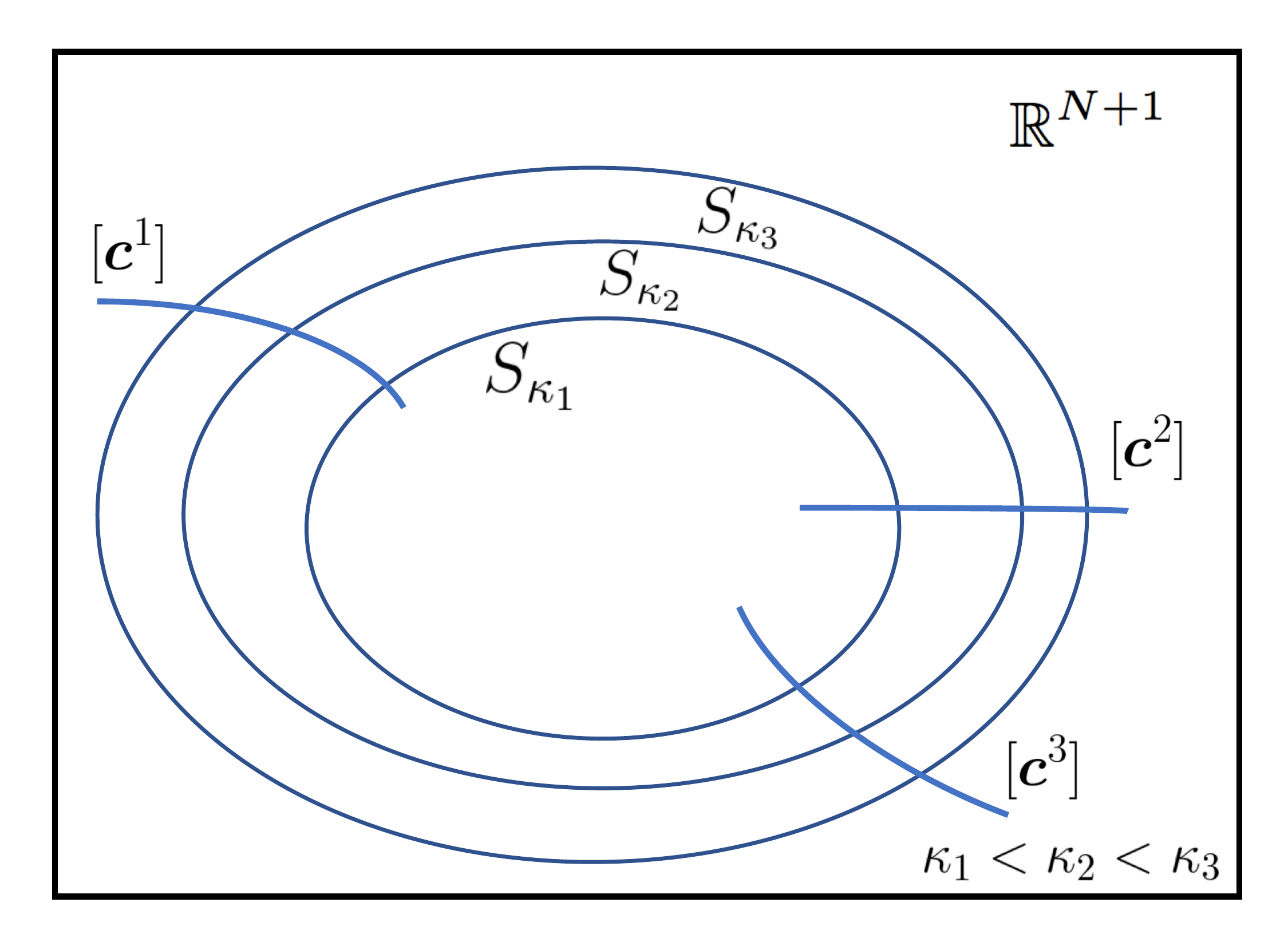}&
\includegraphics[height=1.6 in]{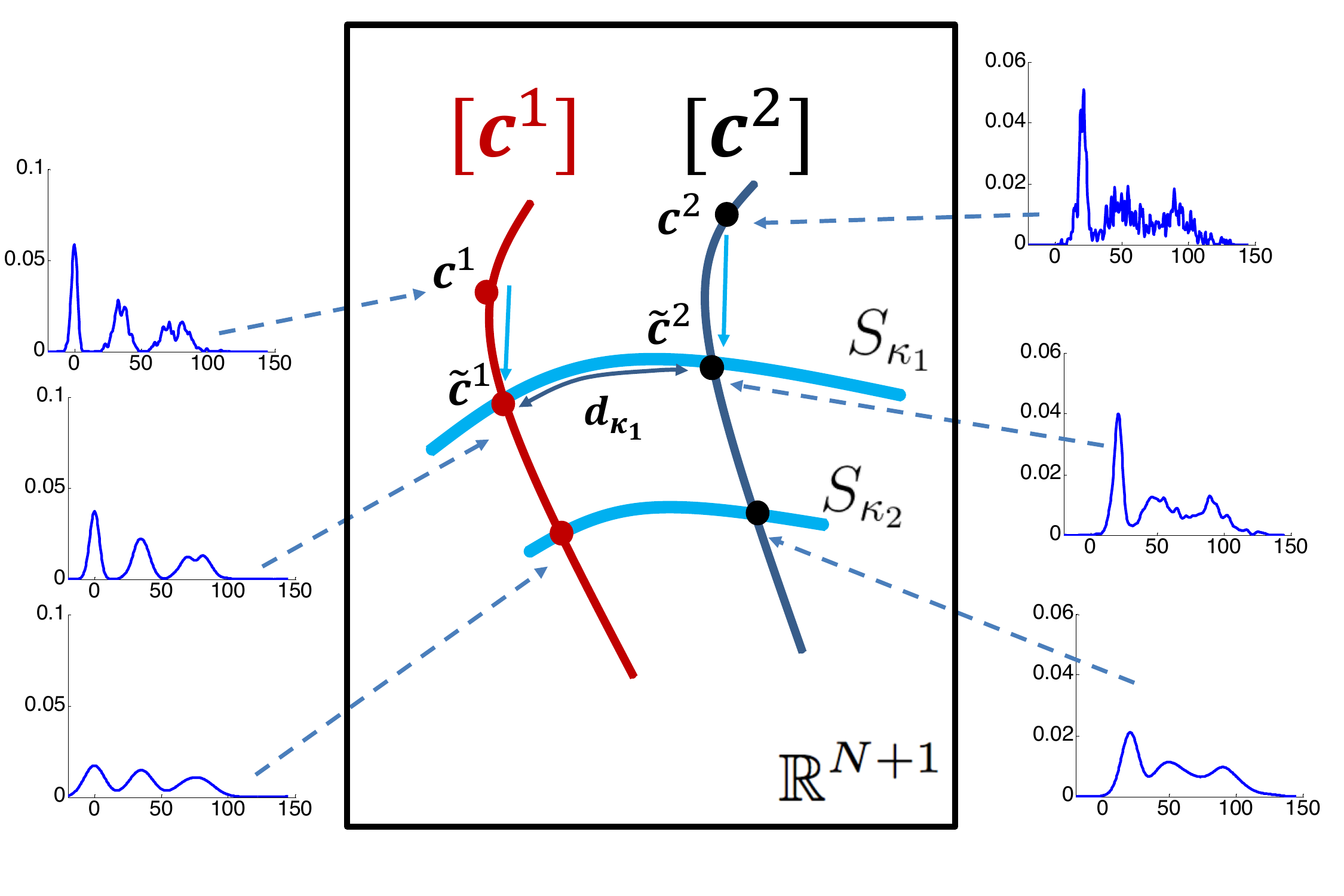} \\
(a) & (b)
\end{tabular}
\caption{ (a) Cartoon illustration of geometry of the representation space $\real^{N+1}$. $[\bc^i]$'s represent 
radial orbits, and $S_\kappa$'s represent the ellipsoidal level sets defined in Eqn. (\ref{eqn:section}). (b) Illustration of calculating $d_\kappa$ using Algorithm 1. $\bc^1$ and $\bc^2$ are initially estimated densities. $\tilde{\bc}^1$ and $\tilde{\bc}^2$ in set $S_{\kappa_1}$ have the same smoothness level $\kappa_1$. $d_{\kappa_1}$ denotes the geodesic distance between $\tilde{\bc}^1$ and $\tilde{\bc}^2$. We also can select another smoothness level $\kappa_2$ to calculate their distance $d_{\kappa_2}$.} 
\label{fig:orbit_section}
\end{center}
\end{figure}

We construct an orthogonal section $S$ as follows. First we define a functional $G: \mathcal{F} \rightarrow \real$ by $G(f_0) = \int_\D f_0(x)(L \cdot f_0)(x) dx$. Using the integration by parts, $G$ can be rewritten as $G(f_0) = \int_\D \inner{\nabla f_0(x)} {\nabla f_0(x) }dx$.
Since $G$ relates to the norm of the gradient, it measures 
the {\it first order roughness} of function $f_0$. 
Also, since $f_0$ is represented by its coefficients as 
an element of $\real^{N+1}$, it is convenient to rewrite $G$ as the mapping 
$G: \real^{N+1} \rightarrow \real$ given by $G(\bc) = \sum_{n=0}^N \lambda_n(c_n)^2$. In our paper, $\lambda_0 = 0$ (because $\phi_0$ is a constant, see Section 2.1), so the summation starts from $n=1$. For a positive real constant $\kappa>0$, we define a section $S_\kappa$ under the blurring group $\real$ as 
\begin{equation} \label{eqn:section}
S_\kappa = G^{-1}(\kappa) \in \real^N =\{\bc \in \real^N| \sum_{n=1}^N \lambda_n (c_n)^2 = \kappa, \kappa >0\} \quad.
\end{equation}
Each point in $S_\kappa$ represents a {\it pdf} with smoothness level equal to $\kappa$ (as measured by the $G$ function). 
By definition, $S_\kappa$ is a set perpendicular to every orbit and, 
therefore, one can think  of $S_\kappa$ as a level set containing {\it pdf}s at the same level of smoothness.   {A formal proof is presented in the Appendix.} 
Since the $\lambda_n$s are all positive, $S_\kappa$ is actually an $(N-1)$-dimension ellipsoid in $\real^N$. A cartoon illustration of the orbit $[\bc]$ and level set $S_\kappa$ are shown in Figure \ref{fig:orbit_section} panel (a). 

To help understand
these abstract concepts, we use a concrete example. From a random sample $\{x_1,x_2,...,x_T\} (x_i \in \D$, drawn from a density function $f$), 
we construct two estimates,  using the Gaussian kernel in $\D$ and different bandwidths $h_1,h_2$, 
denoted as $\hat{f}_{h_1}$ and $\hat{f}_{h_2}$. We use their finite representations $\bc^1, \bc^2 \in \real^{N+1}$ for analysis ($\Pi(\hat{f}_{h_1}) = \bc^1$ and $\Pi(\hat{f}_{h_2}) = \bc^2$), and let $G(\bc^1)=\kappa_1$ and $G(\bc^2) = \kappa_2$. $\hat{f}_{h_2}(x)$ lies in the same orbit as $\hat{f}_{h_1}$, but has a different smoothness level ($\kappa_2 \neq \kappa_1$ if $h_1 \neq h_2$). 
Now we want to bring $\hat{f}_{h_1}$ and $\hat{f}_{h_2}$ to the same smoothness level. Without loss of generality, let us assume $h_2>h_1$.
In this case we smooth $\hat{f}_{h_1}$ (i.e., increase $h_1$) to increase its smoothness level to $\kappa_2$. The precise amount of smoothing required 
can be solved by finding a $t^* \in \real$ such that:

$$G((t^*,\bc^1)) = \sum_{n=1}^N \lambda_n e^{-2\lambda_n t^*} (c_n^1)^2 = \kappa_2, \text{     where } \sum_{n=1}^N \lambda_n (c_n^1)^2 = \kappa_1 .$$
This is the same as finding the intersection of the orbit $[\bc^1]$ with the level set $S_{\kappa_2}$. Due to the monotonicity of the equation with respect to the parameter $t$, we can use the bisection method to solve for $t^*$. If we have more than two {\it pdf} 
estimates at different smoothness levels, we can always choose a certain smoothness level, say $\kappa$, and bring them all to this level.

\subsection{Measure Difference using Geodesic Distance}

 The next problem is how to quantify the difference between the two estimates after we bring them to the same section. An idea is to use the $\ltwo$ distance: $d(\tilde{f}_1,\tilde{f}_2) = (\int_\D |\tilde{f}_1(x) - \tilde{f}_2(x)|^2 dx)^{1/2}$.  
 However, note that $S_\kappa$ has an ellipsoidal structure in terms of the coefficient  vector $\bc \in \real^{N+1}$. A natural way is to treat $S_\kappa$ as a manifold, and quantify differences between points using geodesic
distances. Although it is possible to have analytical expressions for geodesics on ellipsoids in low dimensions, these formulas get very complicated as the dimension grows. In this paper, we use a numerical method called {\it path-straightening} algorithm
\citep{klassen-srivastava-3Dcurves-ECCV} to calculate the geodesic distance on an ellipsoid $S_\kappa$, and denote it as $d_\kappa$. Details of  this algorithm are presented in the Appendix. Given a numerical tool to compute these geodesic distances, we can now outline the 
full procedure for comparing any two samples on the domain $\D$.\\

\noindent
{\bf Algorithm 1} (Numerical Calculation of $d_\kappa$): Given any two 
arbitrary kernel estimates $\hat{f}_1, \hat{f}_2$, and a smoothness level $\kappa$, the defined $d_\kappa(\hat{f}_1, \hat{f}_2)$ is calculated in the following way:
\begin{enumerate}
\item  Represent $\hat{f}_1, \hat{f}_2$ using coefficients under the defined orthonormal basis $\{\phi_0,\phi_1,...,\phi_N\}$: $\hat{f}_1 = \sum_{n=0}^Nc^1_n \phi_n(x)$, $\hat{f}_2 = \sum_{n=0}^Nc^2_n \phi_i(x)$, and let $\bc^i \equiv \{c^i_n,n=0,...,N\}, i= 1,2$. The orthonormal basis used is discussed in Section \ref{kernelr1}. 

\item Find $t^*_1,t^*_2$ to bring $\hat{f}_1,\hat{f}_2$  to the set (orthogonal section) $S_\kappa$ by solving equations: 
$$\sum_{n=1}^N \lambda_n e^{-2\lambda_n t^*_1} (c_n^1)^2 = \kappa,\ \ \  \sum_{n=1}^N \lambda_n e^{-2\lambda_n t^*_2} (c_n^2)^2 = \kappa\ .
$$ 
Then, let $\tilde{\bc}^i \equiv \{e^{-\lambda_n t^*_i}c_n^i,n=0,...,N  \},i = 1,2$. 

\item Calculate $d_\kappa(\hat{f}_1, \hat{f}_2)$ on the ellipsoid $S_\kappa$ using path-straightening algorithm between two points $\tilde{\bc}^1$ and $\tilde{\bc}^2$.
\end{enumerate}

Figure \ref{fig:orbit_section} panel (b) illustrates Algorithm 1 in a cartoon form.
It shows two orbits $[\bc^1]$ and $[\bc^2]$ associated with two densities $\hat{f}_1, \hat{f}_2$. It also shows the actual densities at different levels of 
smoothing ($\kappa_1>\kappa_2$), for each orbit.

\section{Kernel Density Estimation and Comparison}
In this section, we present the complete framework for 
kernel density estimation, representation and comparison on a unit sphere $\mathbb{S}^d$, and discuss its extensions to  $\real^n$. 

\subsection{Densities on Domain $\D = \mathbb{S}^d$}
To apply our framework to densities on $\mathbb{S}^d$,  we need a Gaussian distribution that 
can be used as the kernel function to estimate densities.  In this paper, we focus on $d=1$ and $2$ but the construction can be generalized
to any $d$ in principle. 
Our method assumes that the Gaussian kernel used in estimation 
must be a heat kernel, i.e., the kernel itself is the solution of the heat equation. \cite{hartman1974} pointed out that the widely used Fisher distribution which often plays the role of normal distribution on $\mathbb{S}^d$ is not a heat kernel. 
The heat kernel Gaussian distribution on the circle $\mathbb{S}^1$ is given as:
\begin{equation}
f({\theta}; { \mu},h) =  (2\pi)^{-1} \left(1 + 2 \sum_{m=0}^\infty \exp(-m^2h)cos(m (\theta - \mu)) \right),
\end{equation}
where $ \theta \in [-\pi, \pi)$ is a point on $\mathbb{S}^1$ and $\theta=0$ represents the ``north'' pole of $\mathbb{S}^1$,  ${\bf \mu} $ is the center of the distribution and $h$ controls the variation.  We can easily verify that this distribution is a solution of the heat equation. When $d>1$, we have a $d$-sphere $\mathbb{S}^d = \{ {\bm x} \in \real^{d+1}: | {\bm x }| = 1\}$, the Gaussian kernel is defined as:
$$f({\bm x}; {\bm \mu},h) =  A_d^{-1}\sum_{m=0}^\infty N_{dm}\exp[-m(m+d-1)h]P_{dm}(\left< {\bm x},{\bm \mu } \right>),$$
where: 
\begin{itemize}
\item  $A_d$ is the area of the sphere $\mathbb{S}^d$, which equals ${2\pi^{(d+1)/2}}/{\Gamma((d+1)/2)}$, 
\item $m(m+d-1)$, for $m=0,1,..., \infty$, are the eigenvalues of the Laplacian on $\mathbb{S}^d$, 
\item $P_{dm}$ is the Legendre polynomial of order $m$ for $\real^{d+1}$, 
\item  $N_{dm}$ is the number of linearly independent homogeneous spherical harmonics of degree $m$ in $\real^{d+1}$,
 and 
 \item $\left<, \right>$ indicates the inner product. 
 \end{itemize}
 Taking $d = 2$ as one example, we have $N_{dm} = (2m+1)$, the Legendre polynomial can be expressed using Rodrigues' formula: $P_{2m}(x) = \frac{1}{2^m m!} \frac{d^m}{dx^m}\left[ (x^2-1)^m\right]$, and $A_2 = 4\pi$. So the heat kernel normal distribution on a unit $2$-sphere is: 
\begin{eqnarray}\label{eq:sphereke}
f({\bf x}; {\bf \mu},h) = \frac{1}{4\pi} \sum_{m=0}^\infty (2m+1)\exp [-m(m+1)h] P_{2m}(\left< {\bf x},{\bf \mu } \right>).
\end{eqnarray}
To estimate the density from a sample $\{{\bm x}_1,{\bm x}_2,...,{\bm x}_T\}$ on $ \mathbb{S}^2$,  the kernel density estimation is given as $\hat{f}_h({\bm x}) = ({4T\pi})^{-1} \sum_{i=1}^T \sum_{m=0}^\infty (2m+1) \exp[-m(m+1)h] P_{2m}(\left<{\bm x}, {{\bm x}_i}\right>)$, where $h$ is the bandwidth parameter. 

To apply our framework, we need to find an orthonormal basis $\{\phi_0,\phi_1,\phi_2,...\}$ for smooth functions on $\mathbb{S}^n$. 
For this,  we focus on two spheres, with $d = 1$ and $2$. For $\mathbb{S}^1$,  we use the Fourier basis ($\ltwo([-\pi,\pi],\real)$):  $\{ \frac{1}{\sqrt{2\pi}},\frac{\cos {\theta}}{\sqrt{\pi}},$ $ \frac{\sin {\theta}}{\sqrt{\pi}},\frac{\cos{2 \theta}}{\sqrt{\pi}},  \frac{\sin{2 \theta}}{\sqrt{\pi}},...,$ $\frac{\sin {m\theta}}{\sqrt{\pi}}, \frac{\cos {m \theta}}{\sqrt{\pi}}\}$. With the Laplace operator $L$ in domain $\ltwo([-\pi,\pi],\real)$, we have $ L  \phi_n =  \lfloor	(n+1)/2 \rfloor^2 \phi_n $, and thus the eigenvalue of $ \phi_n $ is $\lambda_n =  \lfloor	(n+1)/2 \rfloor^2$. 
For $d =2$, we use the spherical harmonics basis as follows. 
Let $\theta \in [0,\pi]$ and $\varphi \in [0,2\pi)$ be the spherical coordinates (to apply the kernel in Eqn. (\ref{eq:sphereke}), we need to represent both ${\bf x}$ and $\mu$ in 
spherical coordinates). The spherical harmonics basis of degree $l$ and order $m$ is denoted by
$Y_l^m(\theta, \varphi)$, where $m = -l,...,0,...,l$. On the unit sphere $\s^2$, we have $L \cdot Y_l^m(\theta, \varphi) = l(l+1)Y_l^m(\theta, \varphi)$, and thus if we rearrange the spherical harmonics in the order of $\{Y_0^0,Y_1^{-1},Y_1^0,Y_1^1,Y_2^{-2},...\}$, the corresponding eigenvalue $\lambda_n$ are $\{ 0, 2,2,2,6,... \}$. With the bases established, 
each density can now be represented as the linear combination of the basis functions and coefficients, and
Algorithm 1 can be applied to compare densities. 


\subsection{Extension to Euclidean Domains $\D = \real^n$ }
\label{kernelr1}
This method can be easily extended to Euclidean 
domains such as $\real^1$ and $\real^2$ for broader applicability. 
Gaussian kernels satisfying the heat equation are readily available for these domains \citep{lafferty2005diffusion}. 
However, there is a technical issue in that these domains are not compact. Even we restrict to intervals such as $[0,1]$ and $[0,1]^2$, 
there are some technical problems in directly applying the previously developed framework. 
Note that to represent a density function in $\D$ we need an orthonormal Hilbert basis of $\ltwo(\D,\real)$ with the property that $L \cdot \phi_n = \lambda_n \phi_n$, where $\phi_n$ is one of the basis functions. In terms of this basis, the  heat equation can be solved explicitly; by flowing this solution in the time direction, we obtain an $\real$ action that provides a very natural way to ``spread out'' Gaussian type functions. If $\D$ is non-compact or has a boundary, all of this is either impossible, or much more difficult (requiring a choice of boundary conditions).

In this paper, to handle data in domains such as $\D = \real^1$ and $\real^2$, we first apply a state-of-the-art 
kernel density estimator in  the original domain, and then detect the boundaries of the estimated densities. With these boundaries, we map the 
estimated domain to $\s^1$ or $\s^1 \times \s^1$, and thus wrapping the 
estimated density onto these spherical domains. 
To be more specific, for ${\bm x } \in \real^d$, the Gaussian kernel used is $K_h({\bm x})$.  Let $\{{{\bm x}_1,{\bm x}_2,...,{\bm x}_T}\}$ be a sample of $d$-variate random vectors drawn from an unknown distribution with density function $f$. The Gaussian kernel density estimate is $\hat{f}_h( {\bm x}) =T^{-1} \sum_{i=i}^n K_h( {{\bm x}-{\bm x}_i})$.  We then detect the boundary of $\hat{f}_h$ in $\D$, wrap the function to $[-\pi,\pi]^d$, and rescale all estimated densities to this domain. If we have multiple functions, we detect their boundaries simultaneously and select an large interval that encloses all individual boundary as the shared boundary for all functions. According to the final boundary, we wrap all functions to $[-\pi,\pi]^d$ for comparison.  For $d =1$, we use Fourier basis on $\s^1$ to represent functions in this space.

\section{Two-sample Hypothesis Test} \label{sc:bootstrap}

Since $d_\kappa$ measures the difference between estimated densities from two samples, 
it is a natural statistic for a two-sample hypothesis test. Here we develop a formal procedure using $d_\kappa$.

 Let $\hat{f}_1$ and $\hat{f}_2$ be two estimated densities from samples $\{x_1,...,x_{T_1}\}$ and $\{y_1,...,y_{T_2}\}$ in $\D$ using the bandwidth $h_1$ and $h_2$, and let $\bc^j = \{c_i^j, i=1,..., \infty\},j=1,2$ denotes the finite representation for $\hat{f}_j$. The statistic $d_\kappa(\hat{f}_1,\hat{f}_2)$ is calculated by the geodesic length between $\tilde{\bc}^1$ and $\tilde{\bc}^2$ on the section $S_\kappa$, for a chosen 
 $\kappa$. Under the assumptions $T_1,T_2 \rightarrow \infty$, $0<T_1/T_2<\infty $, $h_1 = h_2$ and null hypothesis $H_0$ that $f_1 = f_2$, 
 it is possible to simplify the test statistic $d_\kappa$ by replacing it with the Euclidean distance (chord length) between $\bc^1$ and $\bc^2$: 
$d_\kappa(\hat{f}_1,\hat{f}_2) \approx \sqrt{\sum_{i=1}^\infty(c_i^1-c_i^2)^2}$.  
Using Parseval's identity it becomes
$\sum_{i=1}^\infty(c_i^1-c_i^2)^2 = \int_{\D}(\hat{f}_1 -\hat{f}_2)^2 dx $. For the simplest case, where $\D=\s^1$ and $h_1 = h_2 = 1$, according to \cite{Anderson199441}, the asymptotic expected value and variance of the test statistic $\Gamma = \int_{-\pi}^{\pi} (\hat{f}_1 -\hat{f}_2)^2 dx$ are given by $E_{H_0}(\Gamma) = (T_1^{-1}+T_2^{-1})J_1$ and  $\text{var}_{H_0}(\Gamma) \sim (T_1^{-1}+T_2^{-1})J_2$, where $J_1 = \int K(x)^2 - \int f^2$, $J_2 = \int \int M(x_1,x_2)^2 f(x_1) f(x_2) dx_1 dx_2,$ $f=f_1 = f_2 $, $M(x_1,x_2) = \int \{ K(x-x_1)-f(x)\} \{K(x-x_2)-f(x)\}dx$, and $K(x)$ is the kernel function.

An asymptotic test may be based on the value of $d_\kappa$, by rejecting the null hypothesis if $d_\kappa$ exceeds the appropriate critical point.
However, even if use the aforementioned simplification,  $d_\kappa^2 \approx \Gamma$, the distribution of $d_\kappa^2$ is not clear \citep{Anderson199441}. Furthermore, the explicit asymptotic distribution of the actual
statistic $d_\kappa$ (the arc-length on the ellipsoid) is even harder to obtain. 
Thus, a more practical approach to the perform two-sample test using $d_\kappa$ is to use the bootstrap method. Fixing a value of $\kappa$ for the whole experiment and letting
 $\{x^*_1,...,x^*_{T_1}\}$ and $\{y^*_1,...,y^*_{T_2}\}$ denote independent re-samples drawn randomly with replacement from the pooled sample set $\{x_1,...,x_{T_1},y_1,...,y_{T_2}\}$, the bootstrap approach is as follows:

\begin{enumerate}
\item[i.] Calculate the geodesic distance between kernel estimated densities from the original two samples $\{x_1,...,x_{T_1}\}$ and $\{y_1,...,y_{T_2}\}$ on a chosen section $S_\kappa$, denoted as $d_\kappa^{0}$.

 \item[ii.] Draw bootstrap samples $\{x^*_1,...,x^*_{T_1}\}$ and $\{y^*_1,...,y^*_{T_2}\}$, and calculate the geodesic distance between kernel estimated densities 
 from these samples on the section $S_\kappa$, denoted as $d^b_\kappa$. Repeat this procedure many times and obtain an empirical distribution of
 $d^b_{\kappa}$. 

\item[iii.] Given $0<\alpha<1$ (the significance level), if $P(d_\kappa^{0}>d^b_{\kappa}) \leq \alpha$, we reject the null hypothesis. 
\end{enumerate} 

\section{Selection of Tuning Parameter $\kappa$}
\label{sc:selectkappa}
Given any constant $\kappa>0$, we can construct an orthogonal section $S_\kappa = G^{-1}(\kappa)$, where $\kappa$ denotes the level of smoothness. Thus, it is important for us to choose a proper $\kappa$ and the corresponding  section $S_\kappa$ for comparing densities. 

Let us consider a scenario of comparing two densities $\hat{f}_1$ and $\hat{f}_2$, and their finite representations ${\bc}^1$ and ${ \bc}^2 \in \real^{N +1}$. Let $G(\hat{f}_1) = \kappa_1$ and $G(\hat{f}_2) = \kappa_2$ (assume $\kappa_1 > \kappa_2$). We can choose a $\kappa \in [\kappa_1, \kappa_2]$ or even a value
outside this interval for evaluating their difference. If we choose a $\kappa> \kappa_2$, we need a $t<0$ to bring $\bc ^2$ to $\kappa$, i.e., $G((t,\bc^2)) = \kappa$. However, this process is susceptible to noise because 
the action is given by $(t,\bc^2) = \{c_0^2,e^{-\lambda_1t}c_1^2,...,e^{-\lambda_Nt}c_N^2 \}$. For a negative $t$, this amounts to inflating the coefficients
exponentially. 
Take the basis for $\s^1$ as one example, where we have $\lambda_N = \lfloor (N + 1)/2 \rfloor$, and $e^{-\lambda_Nt}$ can be a large number even for a small $t<0$. 
If there is some noise in $c_N$ (which is hard to avoid in real data due to numerical errors), such noise will be amplified after multiplying  $e^{-\lambda_Nt}$. Actually, this process is called {\it deblurring} in image processing \citep{liu2014blind}.  Keeping this principle in mind, we propose the following strategies to selected $\kappa$ for the two focused applications:

\begin{enumerate}
\item {\bf Two-sample hypothesis test}. 
In this case, we have only two samples. We first estimate their nonparametric densities with some initial bandwidths which can be obtained by using one of the automatic bandwidth selection methods. Then, we choose the smaller of two $\kappa$s to be the smoothness level for performing the hypothesis testing. 

\item {\bf Comparison of multiple samples}:  Given a set of samples, we first use one of the automatic bandwidth selection methods to estimate their densities. If training data are available, we will select $\kappa$ by cross validation. If training data are not available, we recommend to choose  $\kappa$ such that the $G$-values (smoothness) of most estimated densities (e.g., $~90\%$) are larger than the selected $\kappa$. 
\end{enumerate}

\section{Experimental Results}
In this section, we demonstrate our approach on some selected domains using both simulated and real data. 

\subsection{Simulated Studies on $ \s^1$ or  $\real^1$}

\begin{figure}
\begin{center}
\setlength\tabcolsep{1pt}
\begin{tabular}{cccc}
\includegraphics[height=0.9in]{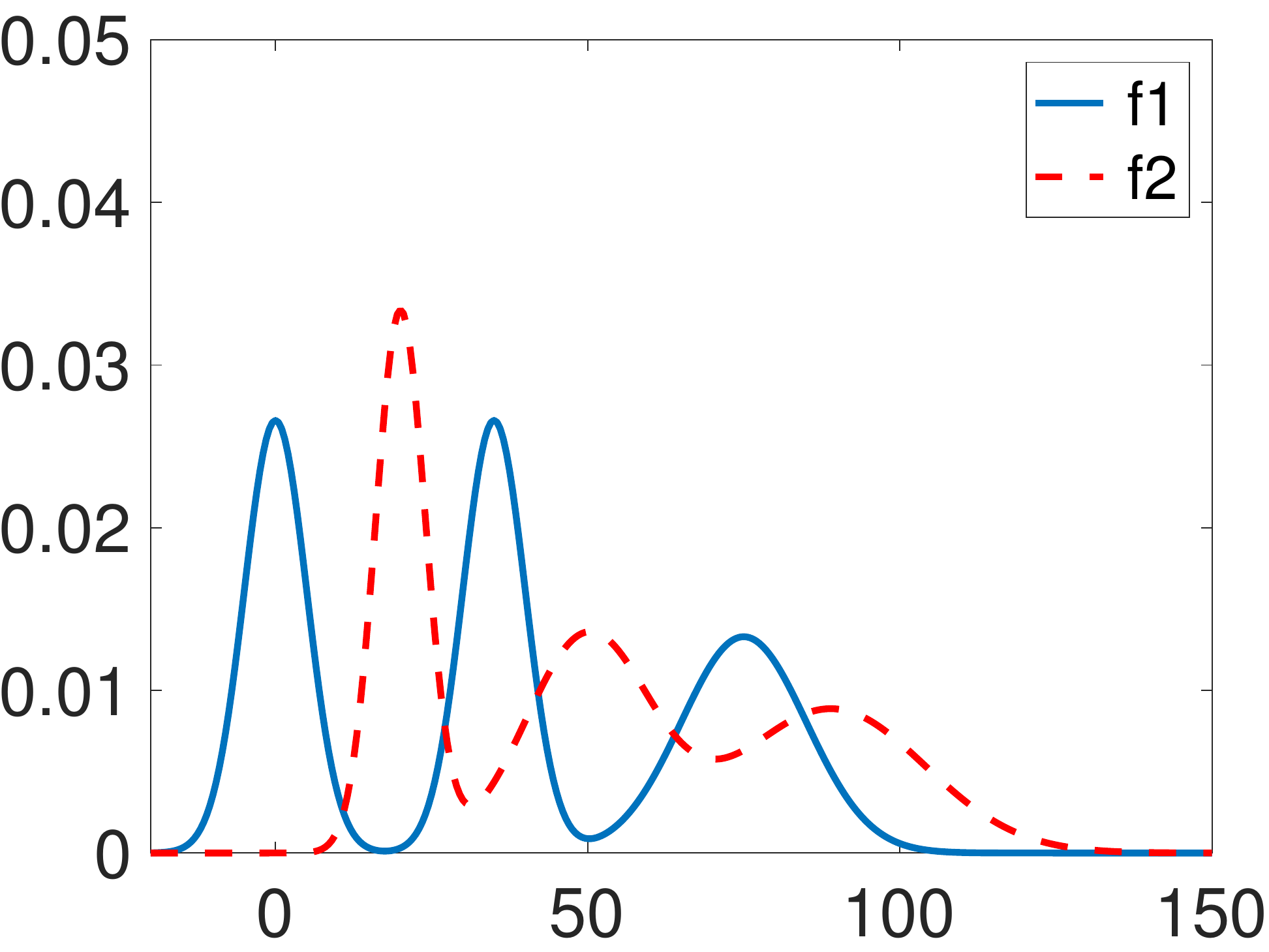}&
\includegraphics[height=0.9in]{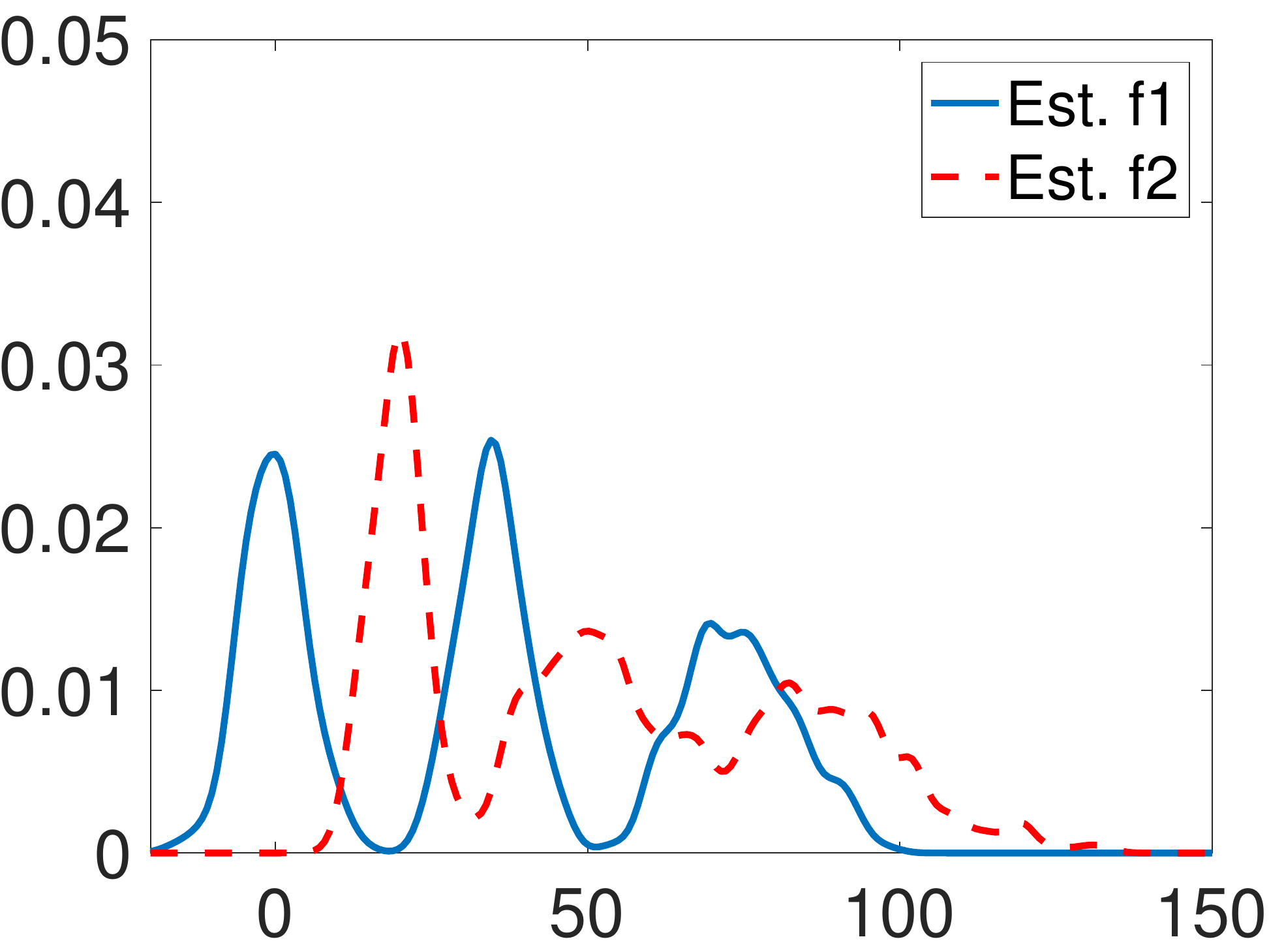}&
\includegraphics[height=0.9in]{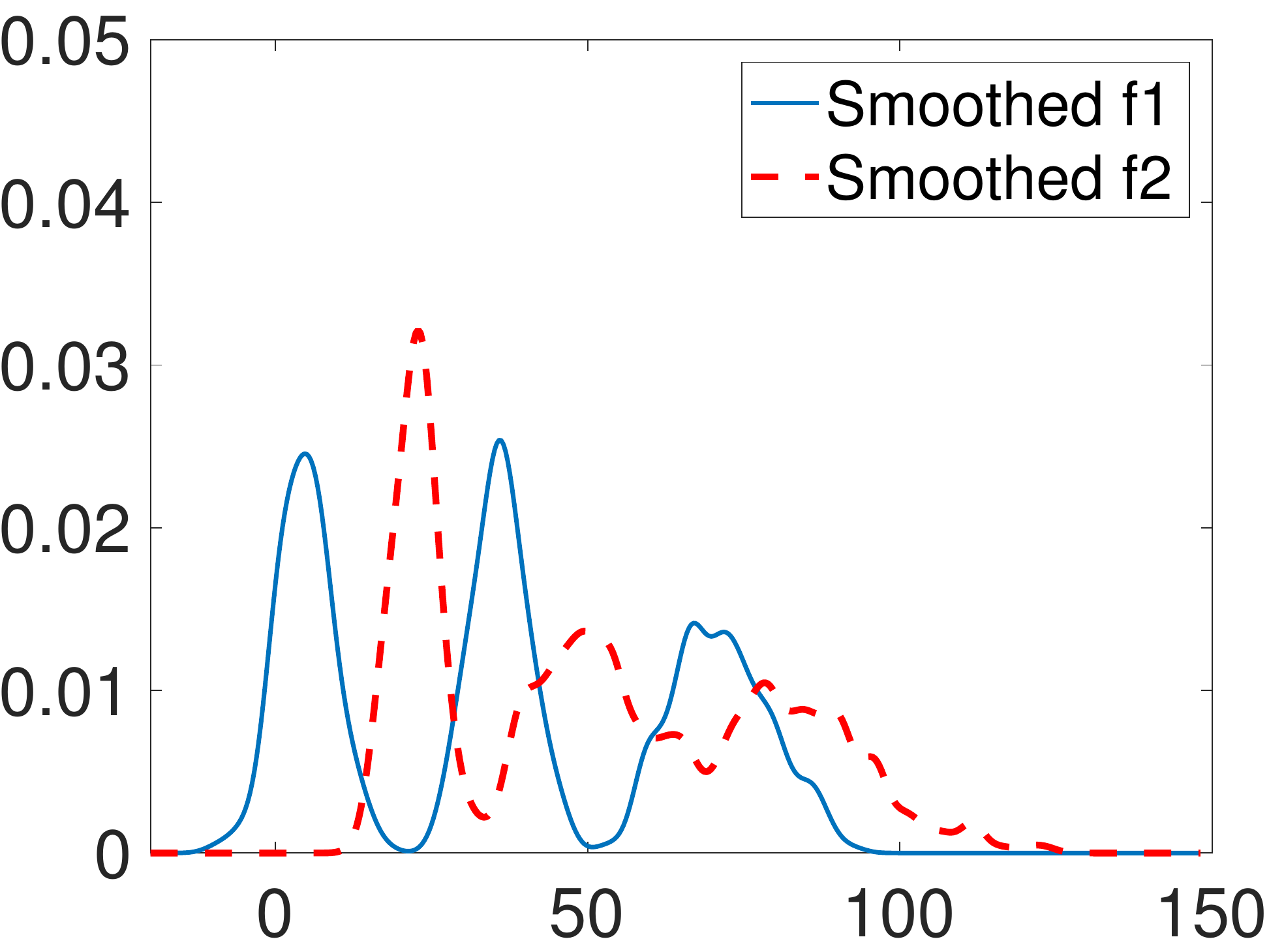} &
\includegraphics[height=0.9in]{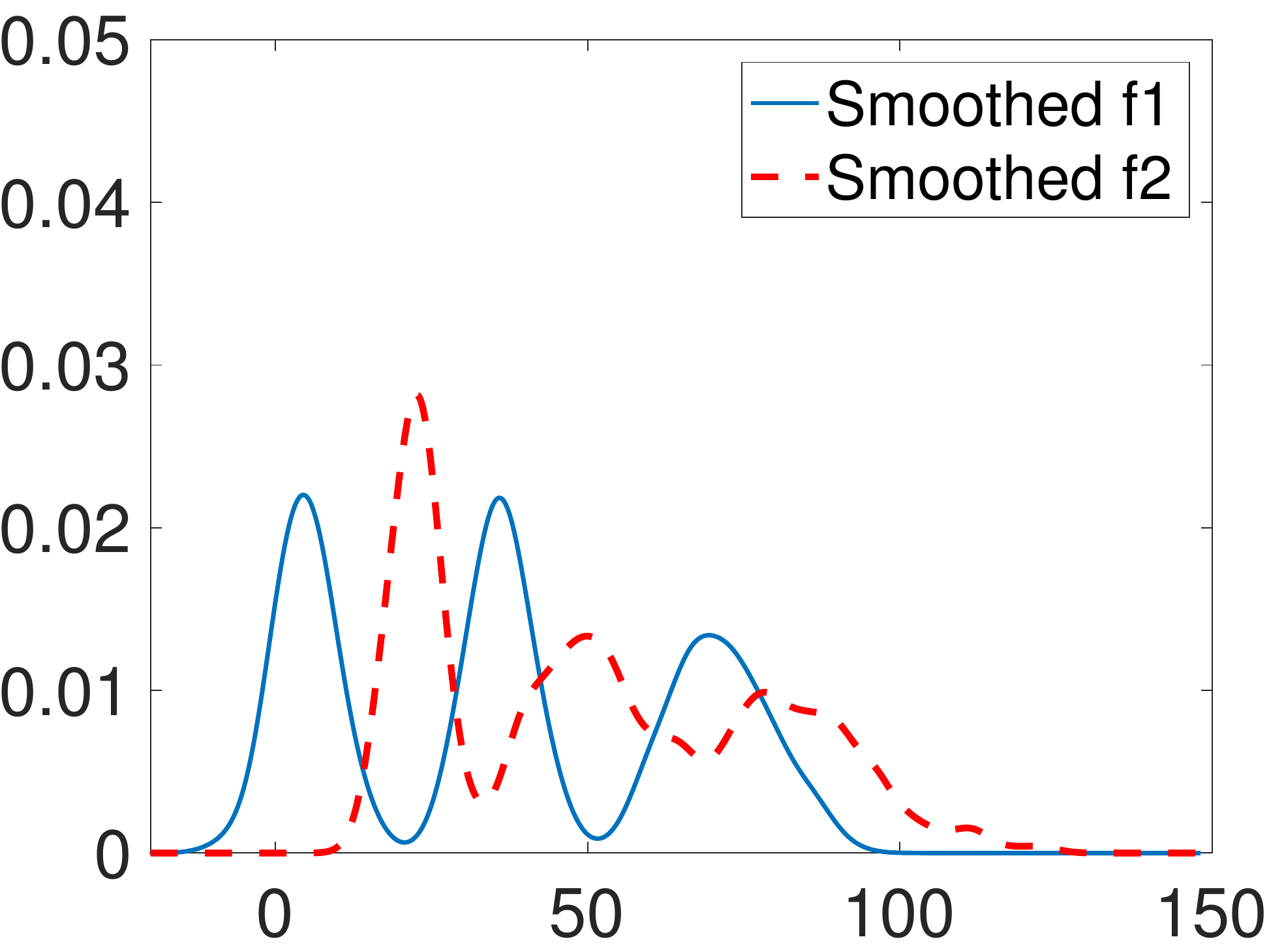} \\
(a) True densities & (b)  Estimated densities & (c)  At $\kappa=7.4$ & (d)  At $\kappa=5$
\end{tabular}
\caption{Illustration of using $d_{\kappa}$ for comparing two kernel estimated densities in $\real^1$. (a) True densities, $f_1$ and $f_2$. (b) Kernel estimated densities from their random samples (each with $600$ samples). (c) and (d) After bringing them to the same smoothness level ($\kappa = 7.4$ and $5$, respectively). Their distances are $d_\kappa = 0.779$  and $d_\kappa = 0.746$.    }
\label{fig:density1}
\end{center}
\end{figure}

\noindent {\bf Comparing Densities Usins $d_{\kappa}$}: 
We first consider the domain $\D=\s^1$.  Densities on $\s^1$ can also be treated as those on an 
interval in $\real^1$ via wrapping $\s^1$ for analysis.  We started from two densities, $f_1$ and $f_2$, 
shown in  Figure \ref{fig:density1} (a). We then sampled $n=600$ points from each and estimated
densities from samples using the kernel method (bandwidths were selected using the method given in \cite{Botev2010}), with estimates 
shown in panel (b).  These functions were wrapped on to the domain $\s^1$ and were represented using basis functions with coefficients. We then manipulated their smoothness levels according to the group action defined in Eqn. (\ref{eqn:groupact}).  Figure \ref{fig:density1}  columns (c) and (d) show the two estimates 
after we matched their smoothness levels to $\kappa = 7.4$ and $\kappa = 5$, respectively. The corresponding geodesic distances 
between these densities are $0.779$ and $0.764$.\\

\begin{figure}
\begin{center}
\begin{tabular}{cc}
\includegraphics[height=1.2in]{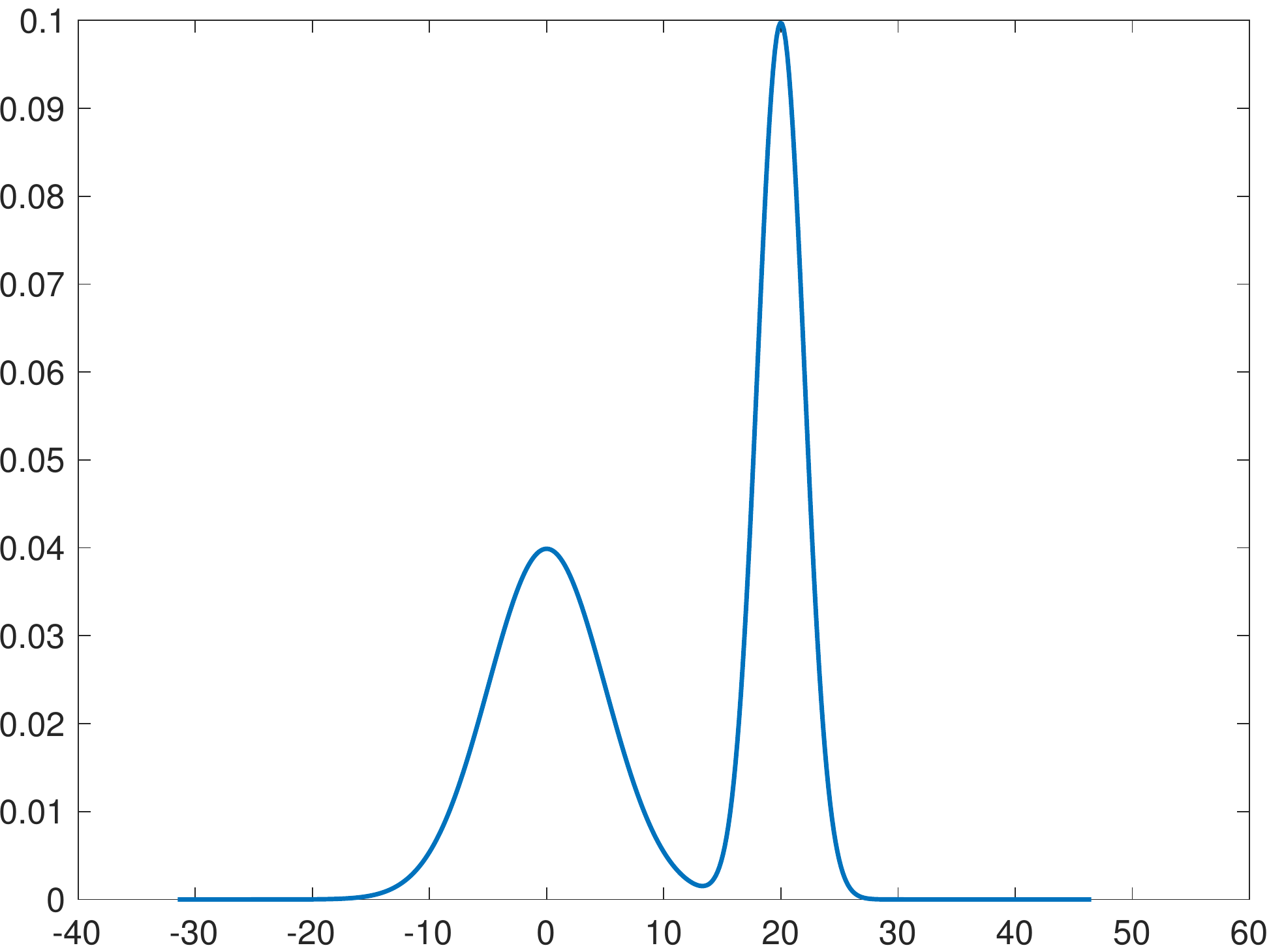}&
\includegraphics[height=1.2in]{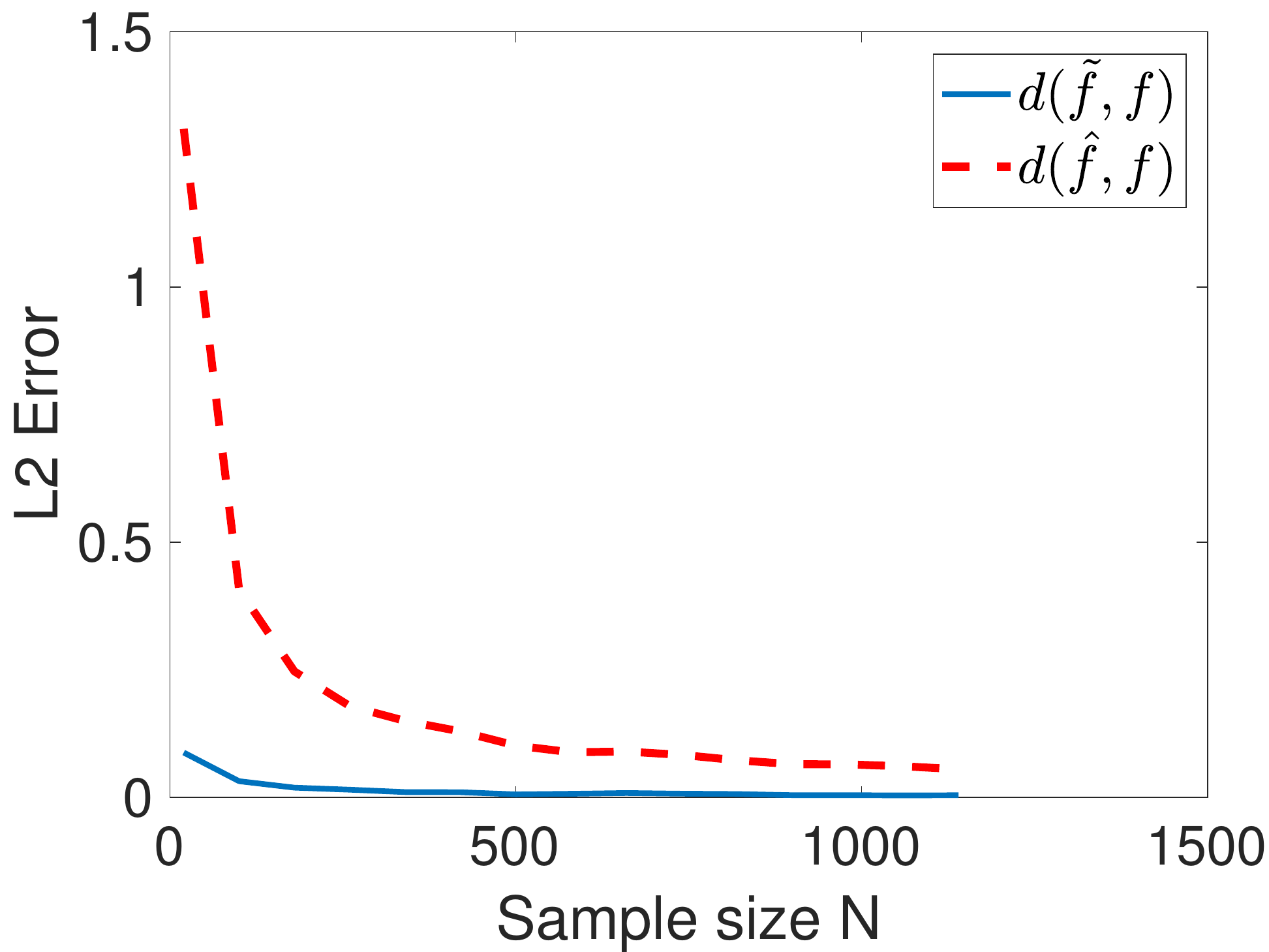}\\
(a) True density & (b) Estimation error
\end{tabular}
\caption{Comparison of density estimation using the optimal bandwidth and optimal smoothness (evaluated by the G-value).  }
\label{fig:optbandwidth}
\end{center}
\end{figure}

\noindent {\bf Utilizing Known Smoothness to Improve Estimatation}: 
In this paper, we have introduced a function $G$ to quantify smoothness of an estimated density function. 
As we know, for kernel density estimates, the bandwidth also controls smoothness of an estimated density. Here we illustrate the connections and differences of 
these two ways of governing smoothness. 
We simulated a density function $f$ and sampled $T$ points from it. Next, we estimated the density in two different 
ways: (a) use the optimal bandwidth that minimizes the asymptotic MISE \citep{scott1987biased} to estimate the density, denoted as $\hat{f}$; 
(b) first estimate the density using a smaller bandwidth and then smooth it to the same smoothness level as that of the underlying true density, denoted as $\tilde{f}$. To compare these two estimates, we used the $\ltwo$ norm to measure the difference between the estimated density and the true density, and the results are presented in  
Figure \ref{fig:optbandwidth}. One can see that  the $\ltwo$ differences between the estimated density and the true density in both methods converge to zero when the sample size increases. However, the latter solution has smaller error and it converges at a faster rate. 
Notably, for small sample sizes, one still can get a very good estimate after bringing the estimated function to the correct smoothness level. 
{The reason is that the smoothness (as quantified by the function $G$) is an intrinsic property of a density function. If any 
prior information about the smoothness of the true density is available, the proposed framework can more efficiently incorporate this prior into the density estimation.  }\\

\begin{figure}
\begin{center}
\begin{tabular}{cc}
\includegraphics[height=1.5in]{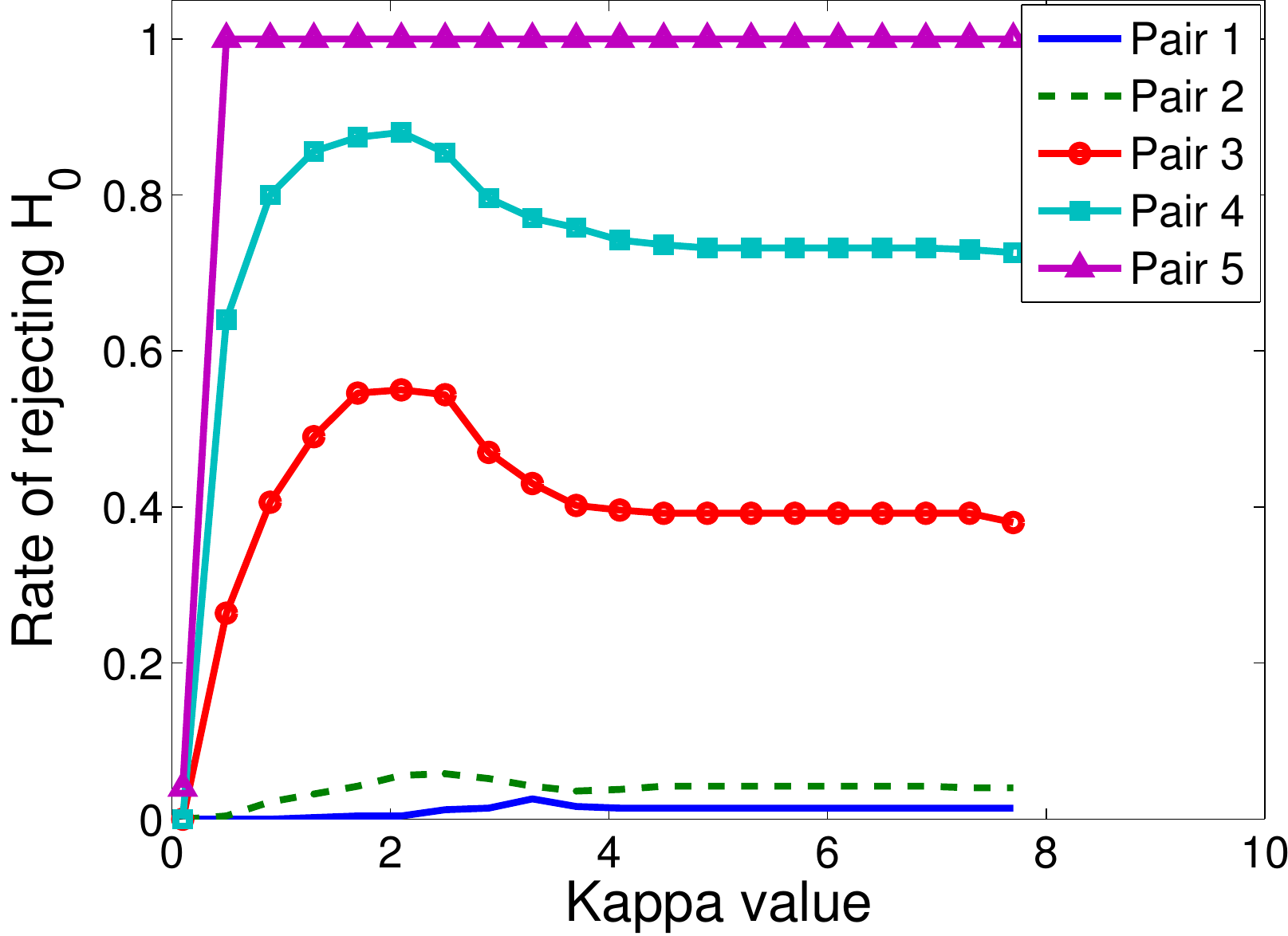}&
\includegraphics[height=1.8in]{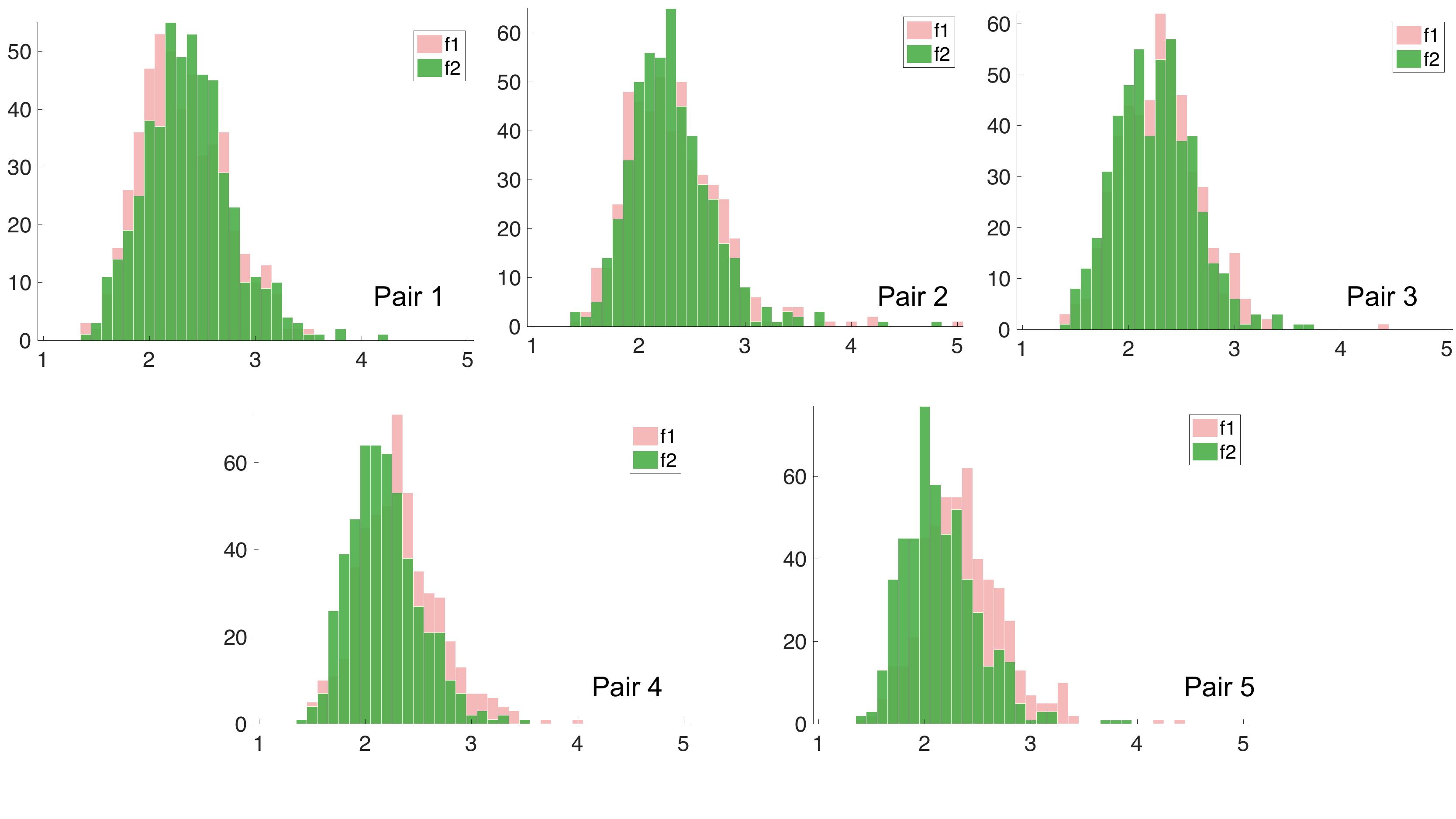}\\
(a) Hypothesis test performance & (b) Histogram of $\kappa$'s   \\
\end{tabular}
\caption{ Effects of $\kappa$ on test performance. (a) $\kappa$ value in x-axis and percentage of rejecting null hypothesis in y-axis. (b) Distributions of $G$-values of estimated densities using \cite{Botev2010} in the simulation. } 
\label{kappainvst}
\end{center}
\end{figure}

\noindent {\bf $d_{\kappa}$ as Test Statistic for Two-sample Test}: 
We are interested in using $d_\kappa$ as a statistic for two-sample hypothesis testing and, further, 
in investigating the effect of $\kappa$ on the power of that test. 
We performed an experiment where we simulated five pairs of densities $f_1$ and $f_2$ with similar smoothness levels, and sampled $n=600$ points from each density. 
The $\mathbb{L}_1$ norm between these five pairs of functions are $0$, $0.06$, $0.14$, $0.17$, and $0.26$, respectively. The initial estimates were
formed using an automatic bandwidth selection method given in \cite{Botev2010}. We then brought them to different pre-specified smoothness levels for 
testing. Figure \ref{kappainvst} (a) shows the results, where the x-axis is the smoothness level $\kappa$, and the y-axis shows the percentage (based on $500$ tests) of rejecting the null hypothesis. From the results, we can see that the selection of $\kappa$ is important. A big $\kappa$ will smooth the estimated densities too much, and therefore, eliminates their difference and reduces the power of the test.  Fortunately, in a relatively large range ($[1.8,2.6]$), we obtain a very good test performance. In panel (b), we show the histogram of smoothness levels of estimated densities (of the 500 runs) using an automatic bandwidth selection method \citep{Botev2010} for each pair of simulated functions. {Following the second procedure of selecting $\kappa$ in Section \ref{sc:selectkappa}, we can select $\kappa \approx 1.8$ for all the five pairs for the hypothesis testing. This $\kappa$ will 
result in good performance: a small type I error (see pair 1), and a good test power (see pairs 2, 3, and 4).} 

\begin{figure}
\begin{center}
\begin{tabular}{cc}
\includegraphics[height=1.2in]{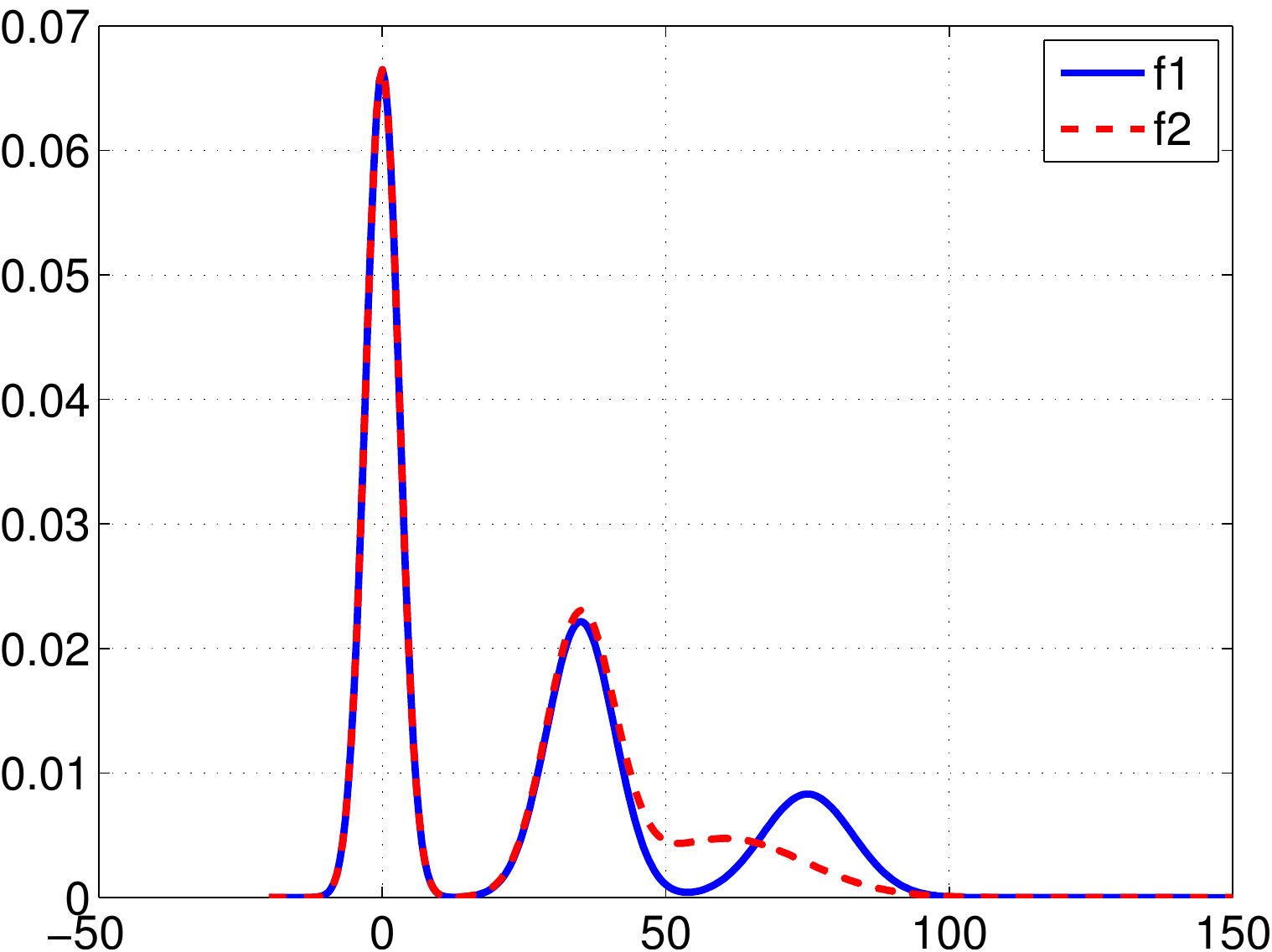}&
\includegraphics[height=1.2in]{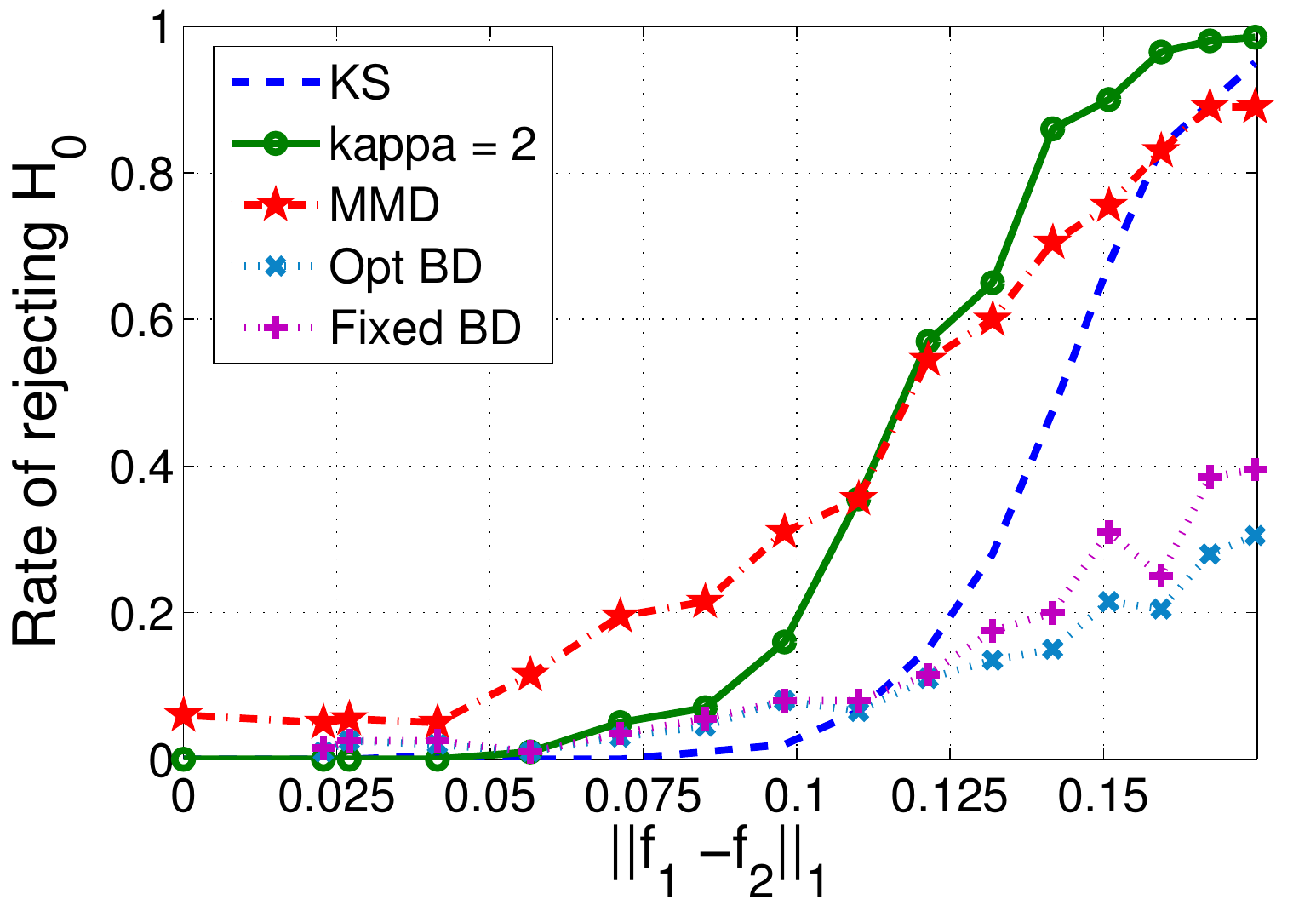}\\
\includegraphics[height=1.2in]{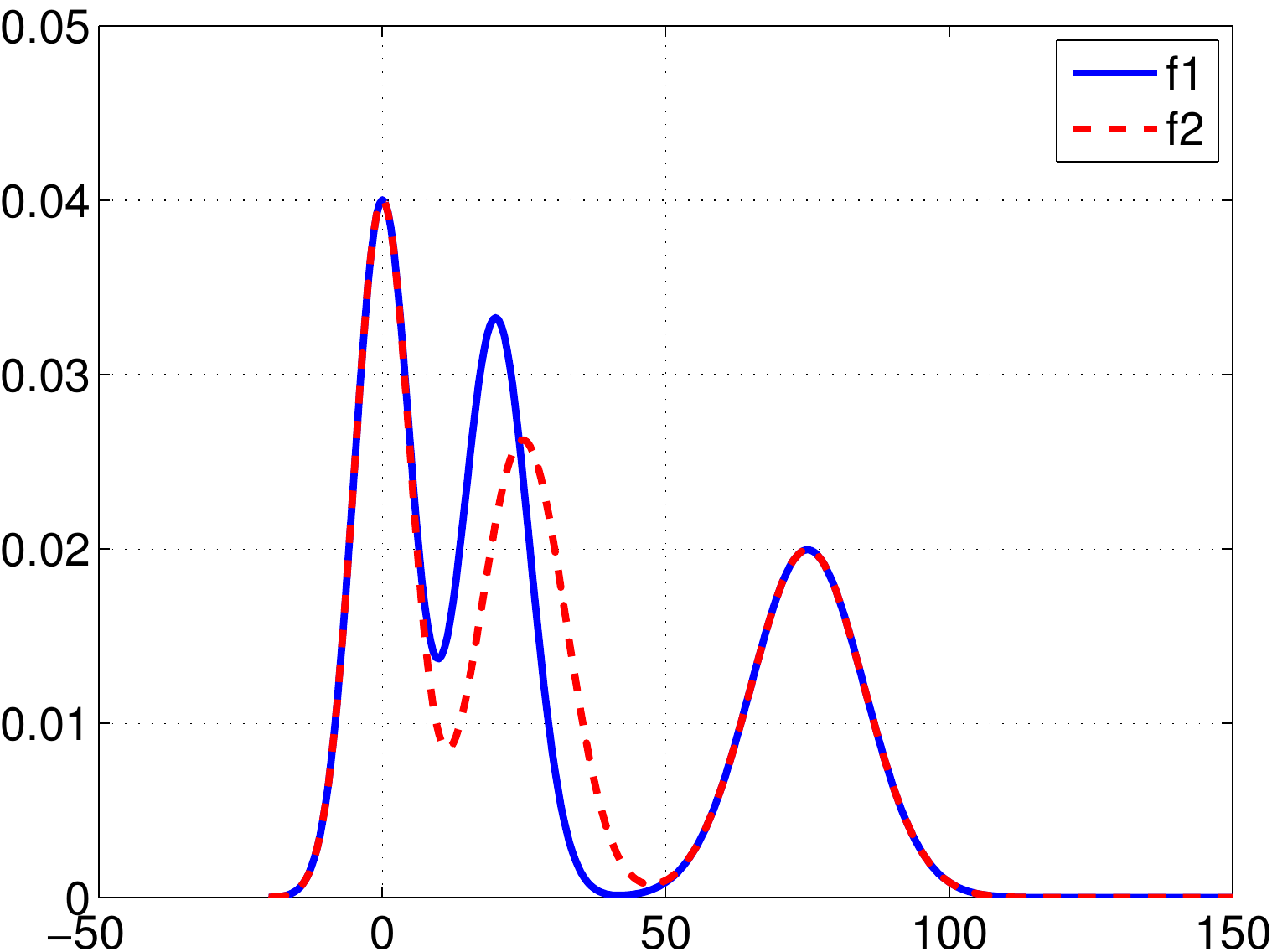}&
\includegraphics[height=1.2in]{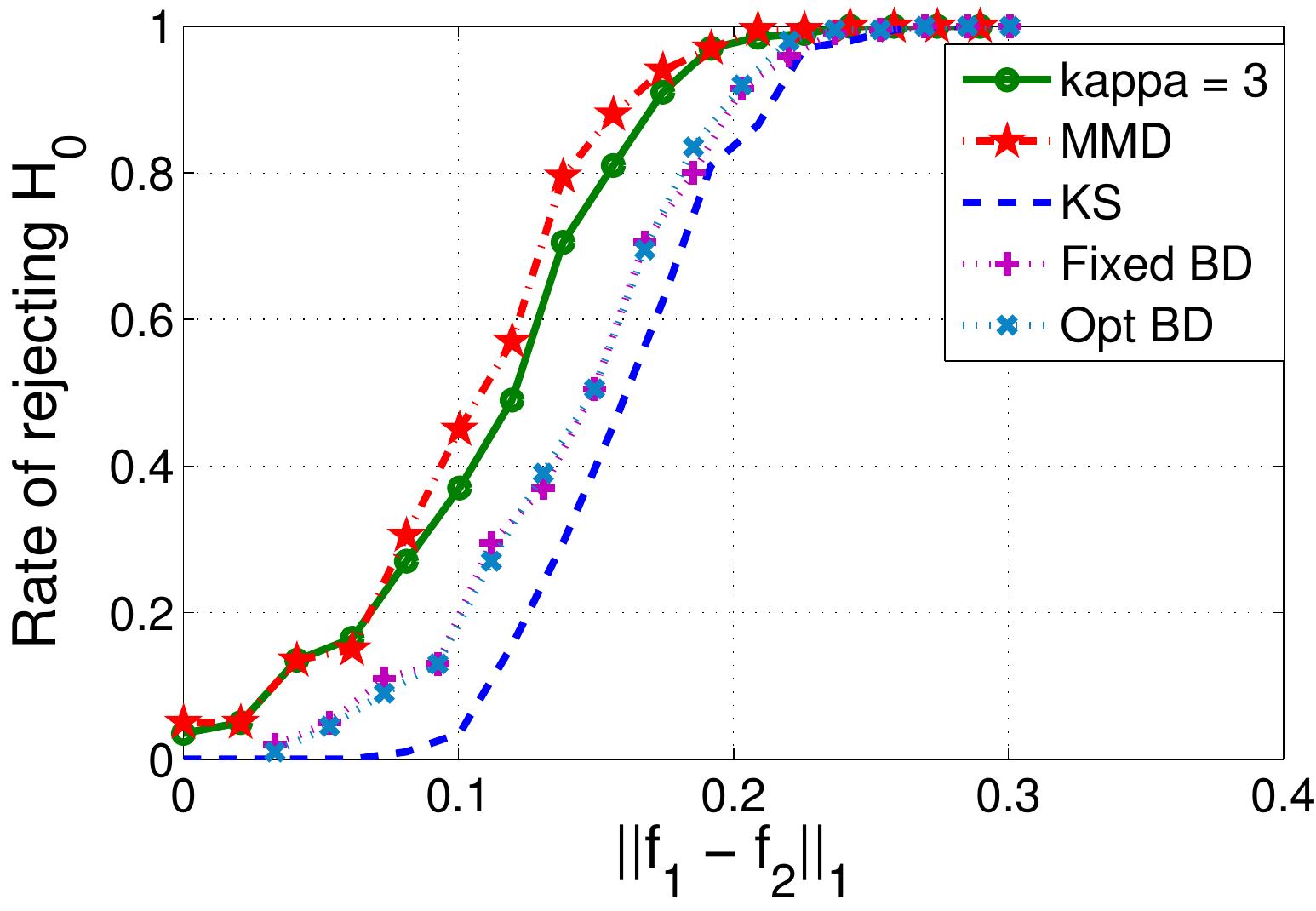}\\
(a) True densities & (b) Test performance \\
\end{tabular}
\caption{Two-sample hypothesis test in $\real^1$. (a) shows one example of the true densities $f_1$ and $f_2$. (b) shows the test performance.} 
\label{fig:hypothesis1}
\end{center}
\end{figure}

We also performed an extensive experiment 
to compare our results with other two-sample test methods. We consider two different scenarios in this experiment: (1) a case where the difference between $f_1$ and $f_2$ 
lies in the tail; and (2) a case where the difference lies in the middle.  Figure \ref{fig:hypothesis1} column (a) shows the simulated densities (the first row shows scenario (1) and the second row shows scenario (2)). We sampled $600$ points from each density, and used them for the testing. We compared  with four other tests: (i) Kolmogorov-Smirnov (KS) test;
(ii) test based on $\ltwo$ distance between estimated densities using a unit bandwidth (Fix BD) \citep{Anderson199441} and (ii) the optimal bandwidth (Opt BD) \citep{Botev2010}; and (iv)  maximum mean discrepancy (MMD) method \citep{Gretton2012,Gretton07} (a Gaussian kernel with a recommended bandwidth by \cite{Gretton2012} was used). The $\kappa$ in our method was chosen to be fixed at $2$.  In scenario (1), our method outperforms  the statistics of KS, Fix BD and Opt BD.  
When the difference between $f_1$ and $f_2$ is small, our method has a smaller chance of rejecting the null hypothesis (a lower type II error)
compared to MMD and, when the difference is large, our method has a bigger chance of rejecting the null hypothesis (a higher test power)
relative to MMD. In scenario (2), the proposed method has a similar test
power with MMD, but outperforms KS, Fix BD and Opt BD.

\subsection{Real Data Application in $\s^1$ or  $\real^1$}

The geodesic distance $d_\kappa$ is not only a statistic for the two-sample hypothesis testing 
but also a metric to quantify differences between non-parametric densities. One potential application is in the computer vision area, where features are extracted and compared using their distributions \citep{liu2003,Chaudhry09hist,Robert2002}, e.g., histograms. However, the choice of the number of bins
has a large influence on the shapes of histograms. Instead of comparing the histograms, a better way of comparing two features is to compare their kernel estimated densities, 
especially when using the distance $d_{\kappa}$.

\begin{figure}
\centering
\begin{tabular}{ccc}
\includegraphics[height=1in]{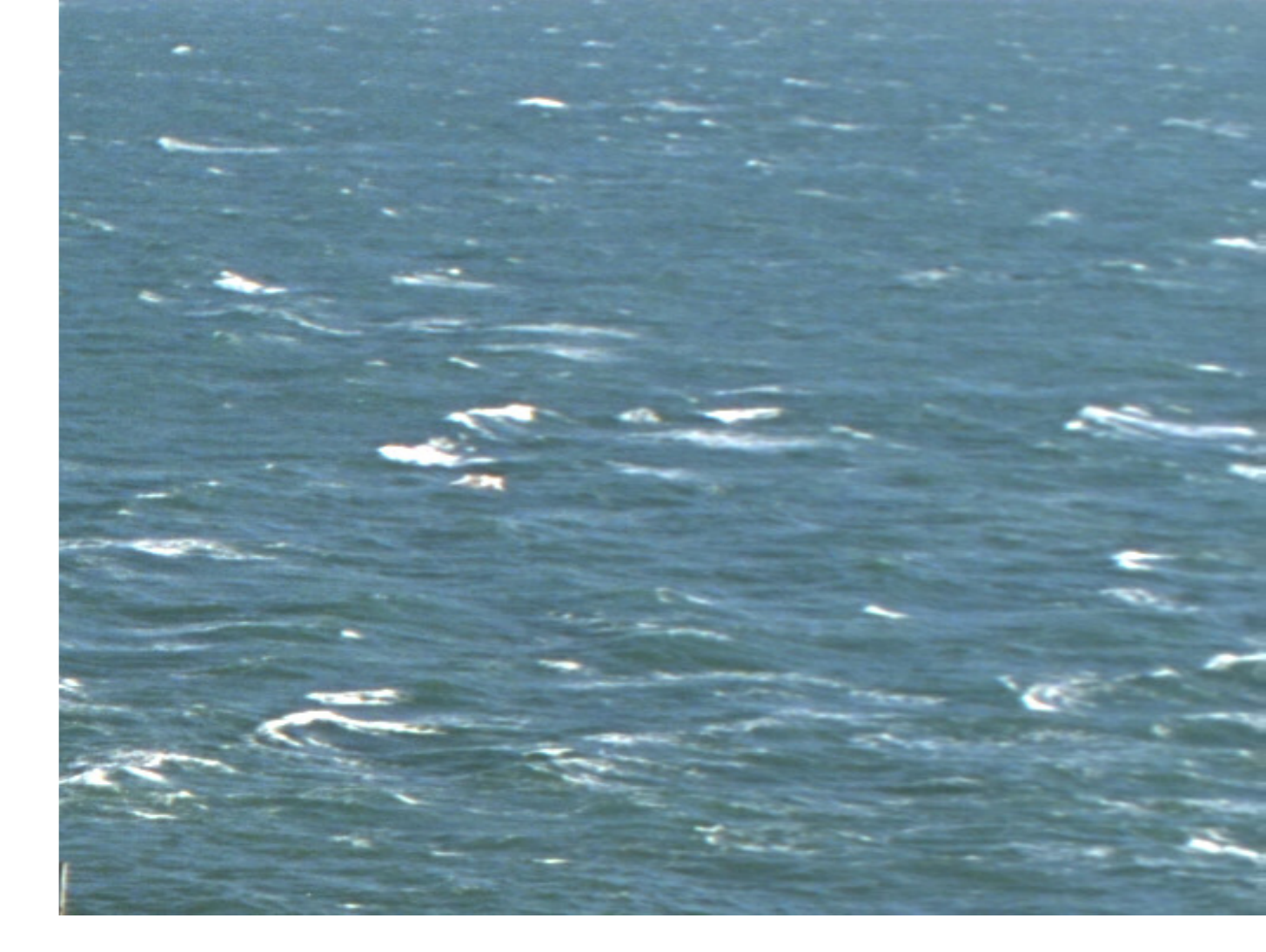}&
\includegraphics[height=1in]{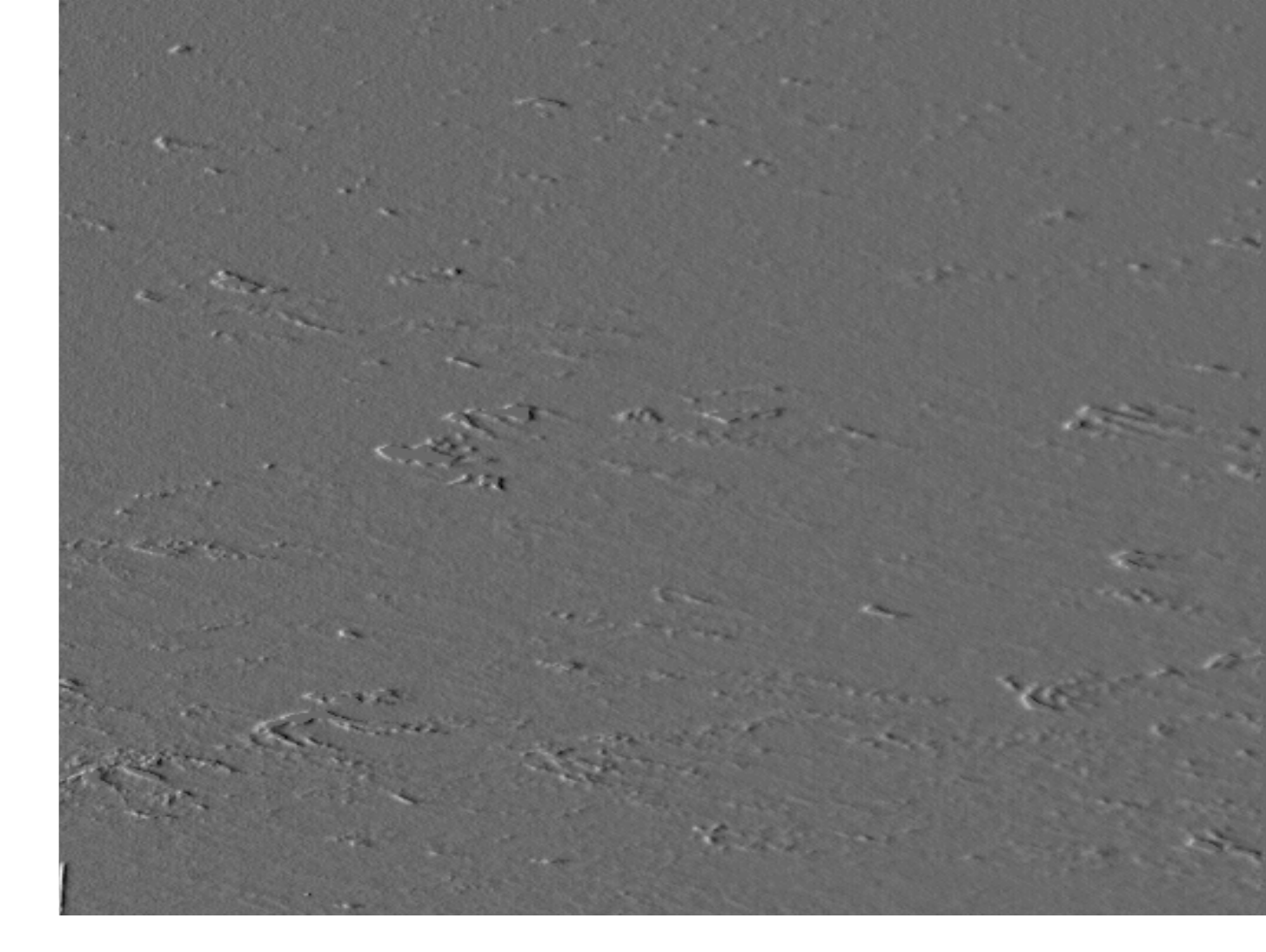}&
\multirow{2}{*}[0.4in]{\includegraphics[height=1.6in]{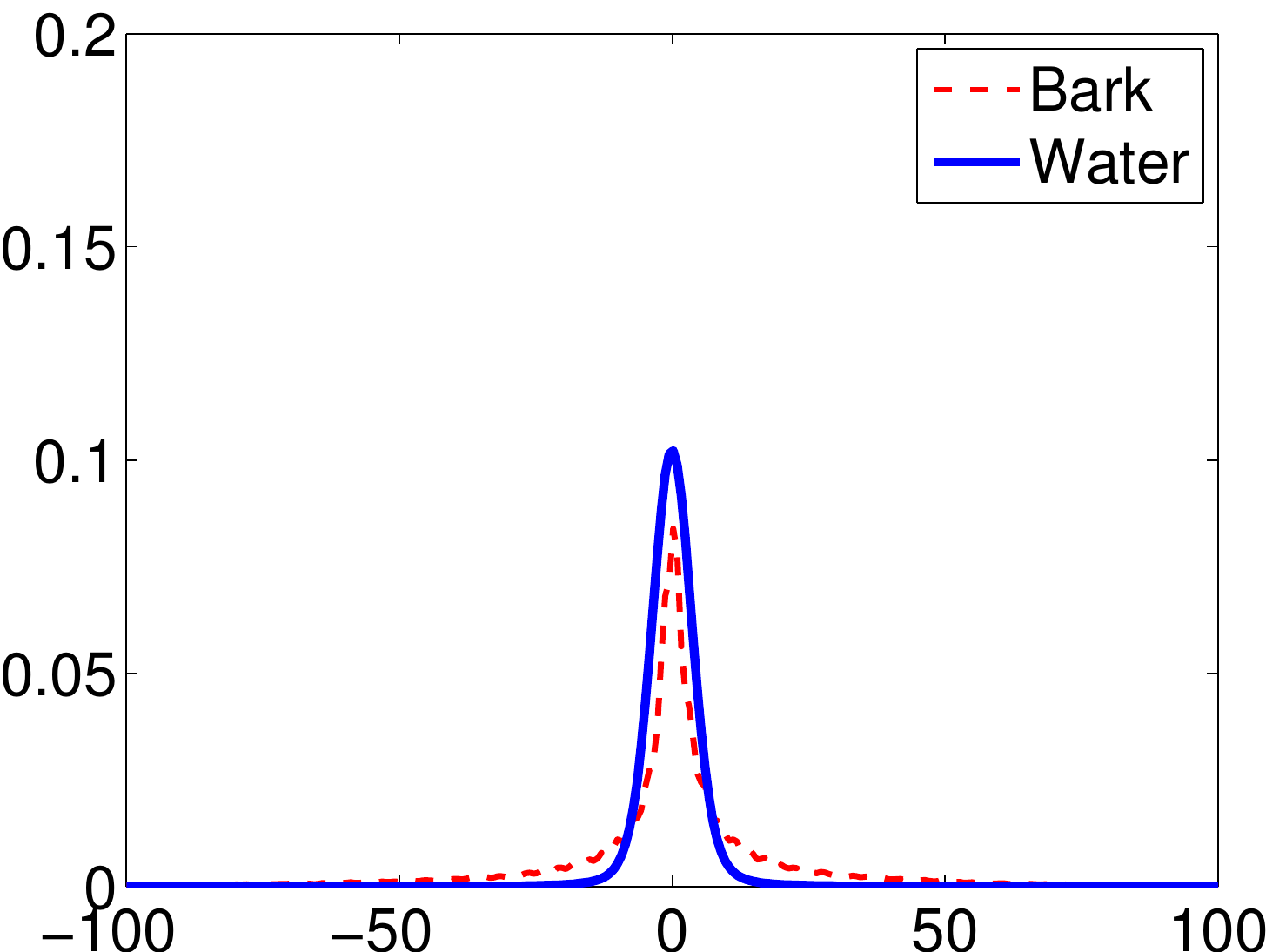}} \\
\includegraphics[height=1in]{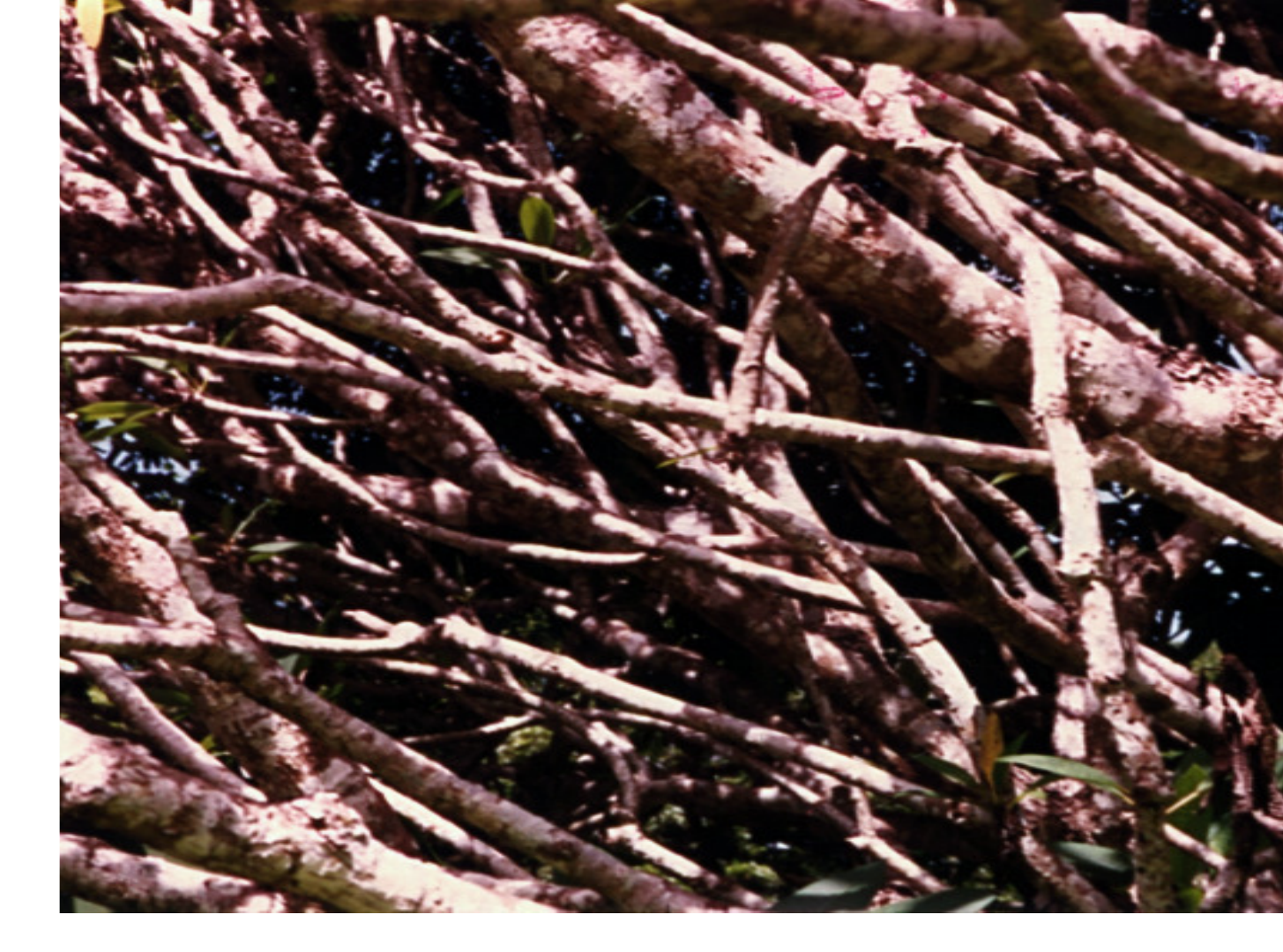}&
\includegraphics[height=1in]{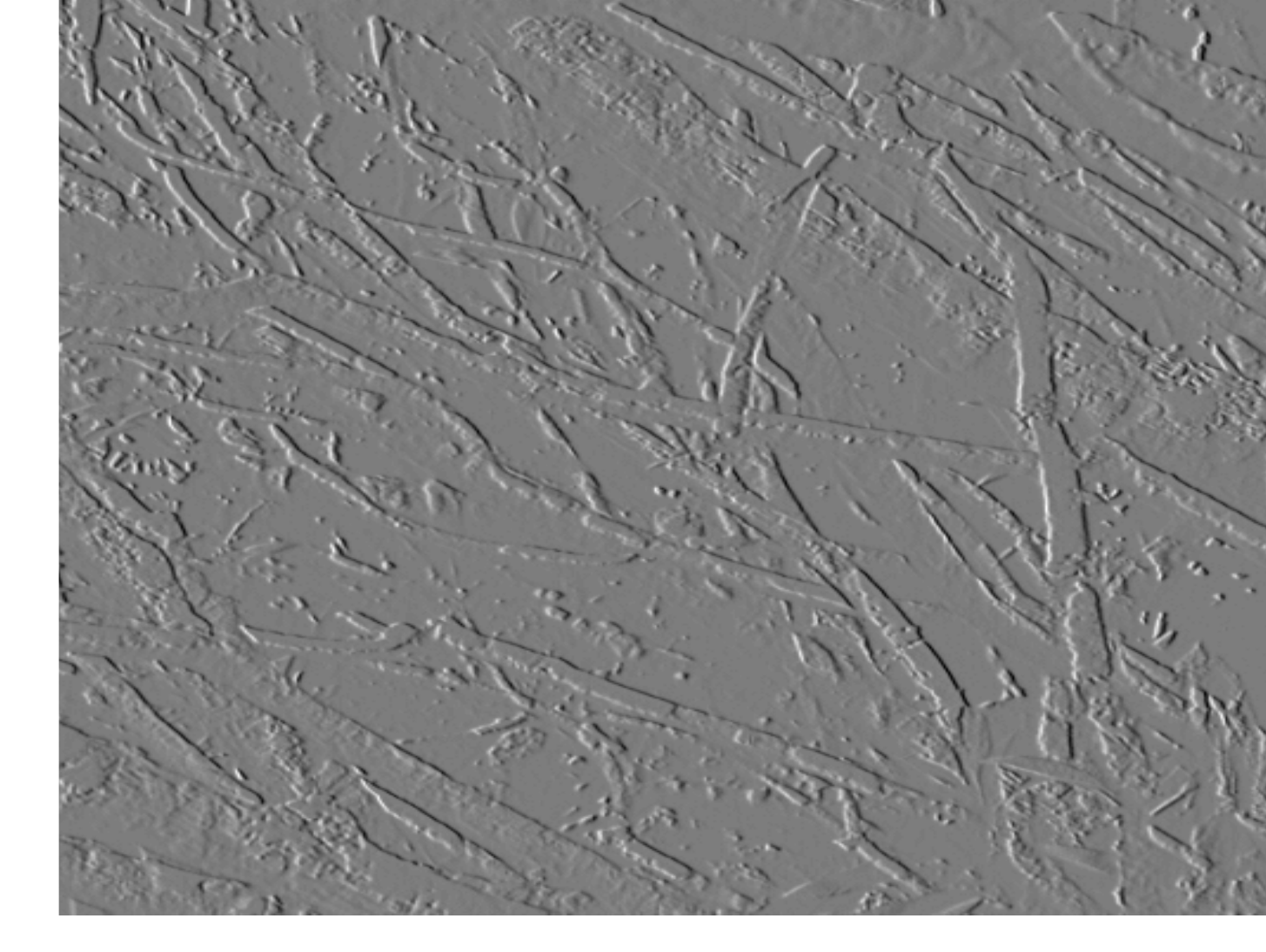}& \\
(a) Original texture images & (b) Features & (c)  Kernel estimated densities\\
\end{tabular}
\caption{Comparison of features using $d_\kappa$. Estimated densities in (c) were brought to the same smoothness level.   }
\label{fig:hist}
\end{figure}

\begin{figure}
\centering
\begin{tabular}{cccc}
\includegraphics[height=0.9in]{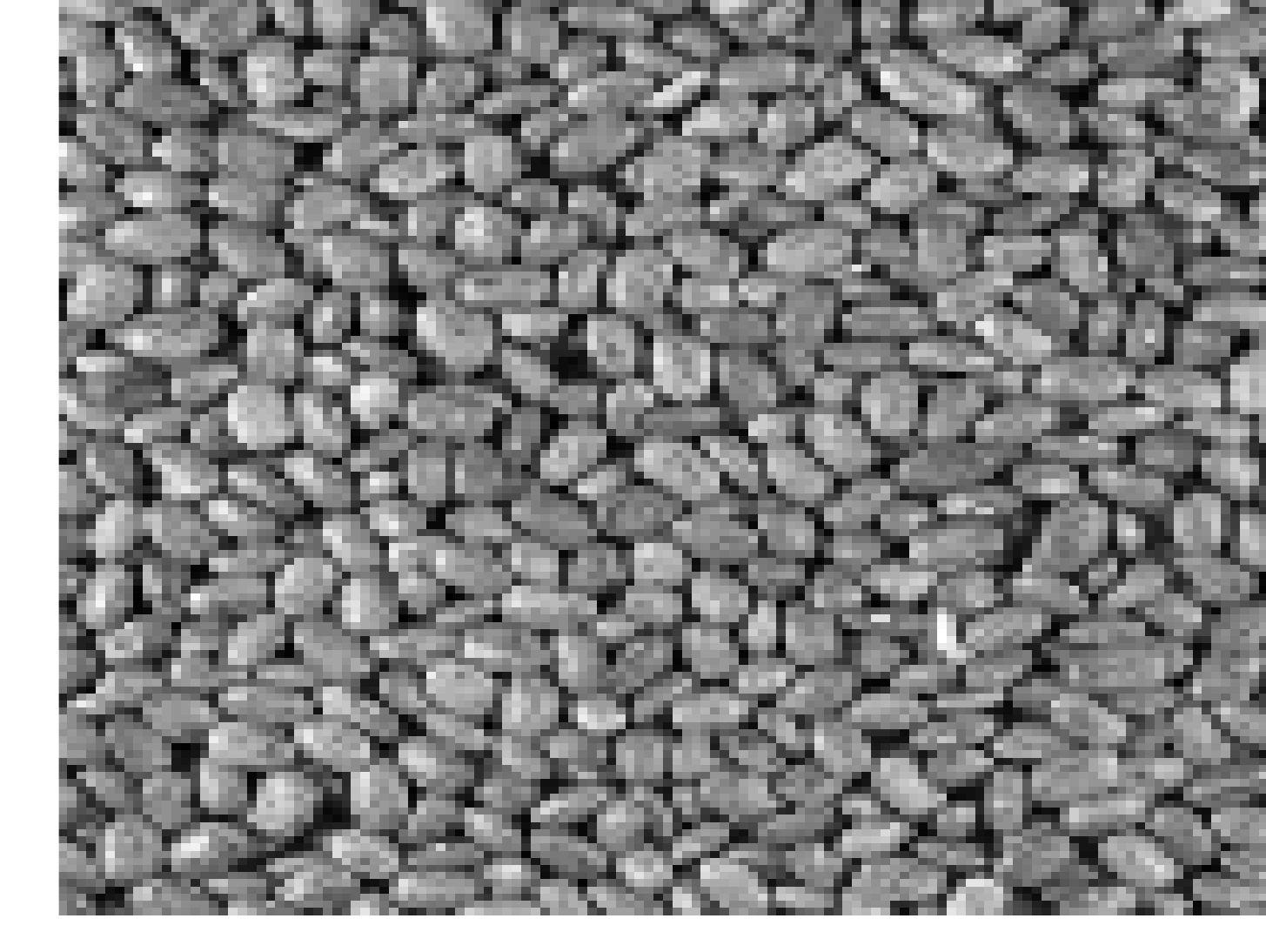}&
\includegraphics[height=0.9in]{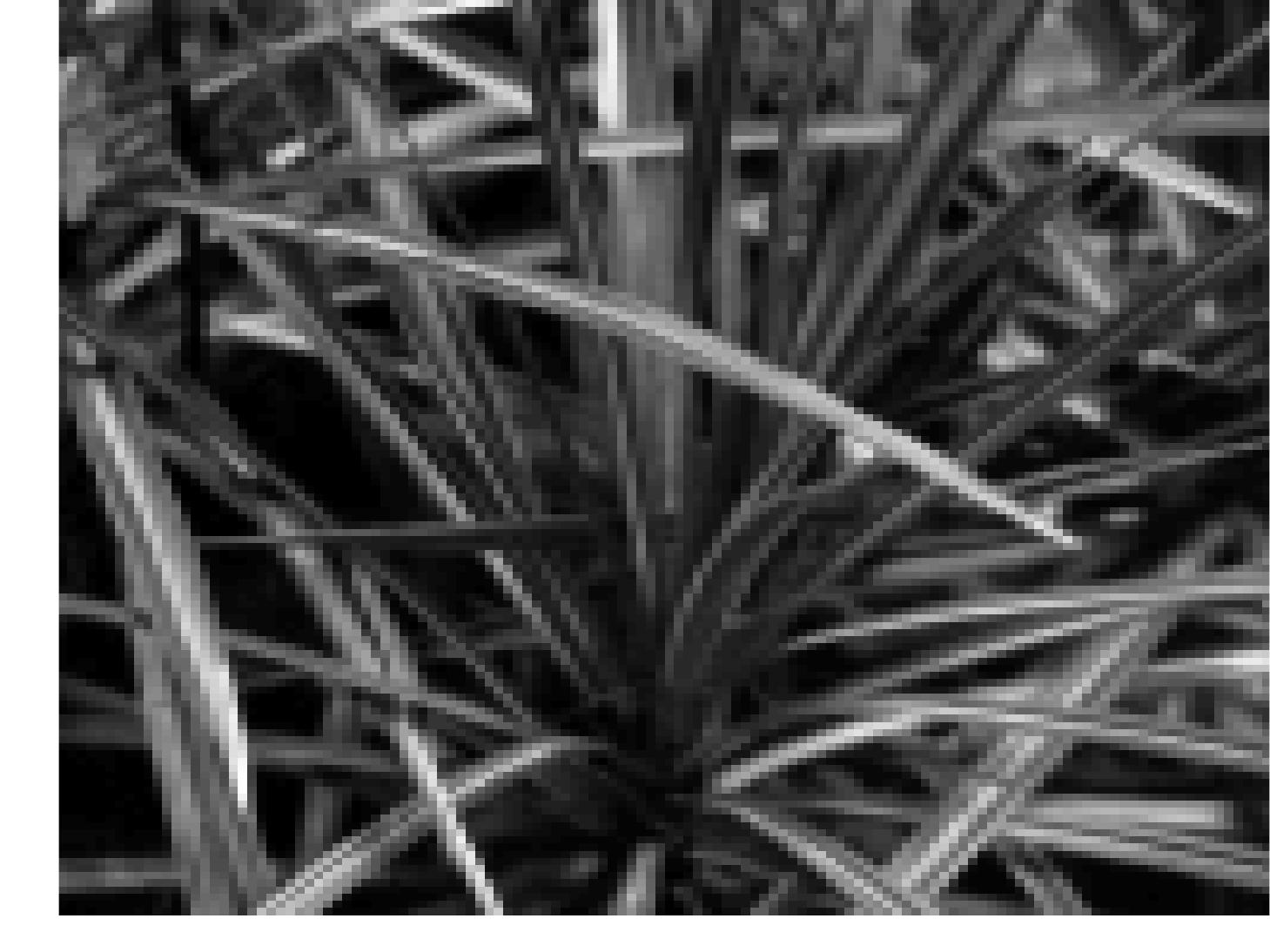}&
\includegraphics[height=0.9in]{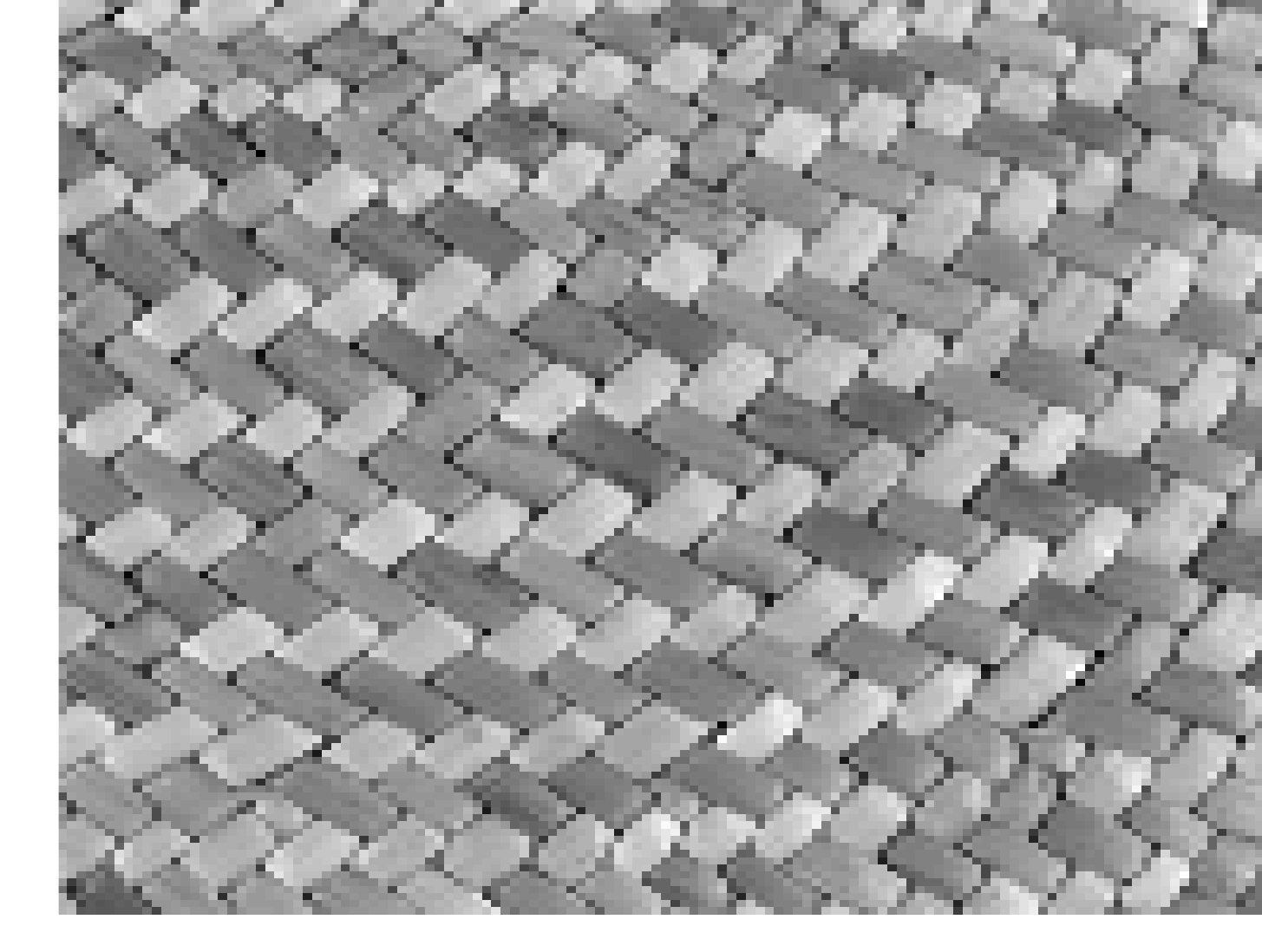}&
\includegraphics[height=0.9in]{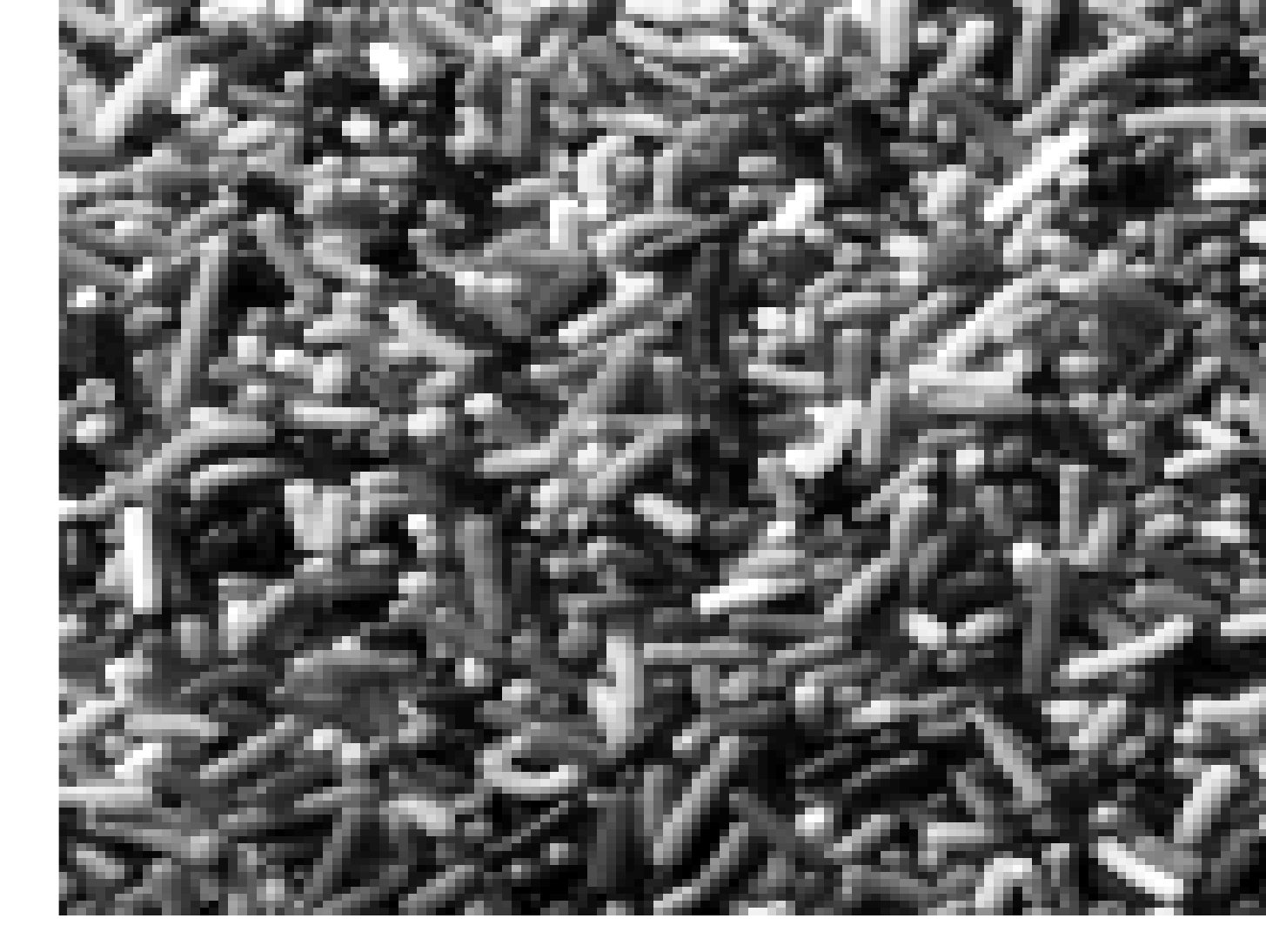}\\
Food & Leaves & Fabric & Food \\
\end{tabular}
\caption{Example images in texture classification dataset.  }
\label{fig:textureexp}
\end{figure}

We performed an experiment involving classification of images 
to illustrate advantages of $d_\kappa$ in comparing image features. According to \cite{liu2003}, the spectral histograms, which are 
nothing but marginal distributions of the image after convolving with some filters, can be used to represent and classify texture images. 
Motivated by this argument,  we convolved each texture image with $6$ Gabor filters \citep{liu2003} of size of $8 \times 8$, and the corresponding kernel estimated marginal distributions were computed and used as features to classify the texture images.  
Figure \ref{fig:hist} illustrates one example of comparing two texture images using $d_\kappa$ ($d_{\kappa=500} = 0.7793$). Our classification dataset contains $54$ texture images from $6$ different categories: leaves, food, fabric, buildings, brick, bark. Each category has $9$ images, and each image has size of $128 \times 128$. 
Some examples of these images are shown in Figure \ref{fig:textureexp}. In the classification experiment, we chose 
these images one-by-one as queries and
found their nearest neighbor under the metric mentioned above. If the nearest neighbor image belongs to the same category as the query image, we consider it as a success retrieval, otherwise as a failed one. We performed this process for every image in the dataset. Table \ref{tab:textureclass} shows the classification result. The numbers in the table are number of successful retrievals for each category. We compared our method with other five similarity measures. Assuming $hist_1$ and $hist_2$ represent two histograms of features, while $\hat{f}_1$ and $\hat{f}_2$ represent the corresponding estimated densities, these similarity measures are:

\begin{enumerate}
\item {  Hist. (20)}: $\ltwo$ distance between histograms with $20$ bins, defined as $\D(hist_1, hist_2) = \| hist_1 - hist_2\|_2$. 
\item { Hist. (100)}: $\ltwo$ distance between histograms with $100$ bins.
\item {$\ltwo$}:  $\ltwo$ distance between estimated densities,  defined as $\D(\hat{f}_1,\hat{f}_2) = ( \int (\hat{f}_1 - \hat{f}_2)^2 dx) ^{1/2}$. 
\item {$\chi^2$}: $\chi^2$ distance,  defined as $\D(\hat{f}_1,\hat{f}_2) = \int{ {(\hat{f}_1 - \hat{f}_2)^2}/{(\hat{f}_1 + \hat{f}_2)}}dx$. 
\item {Bhatt.}: Bhattacharyya distance, defined as $\D(\hat{f}_1,\hat{f}_2) = 1 - \int {(\hat{f}_1 \hat{f}_2)^{1/2}}dx$.
\end{enumerate}

The kernel densities in this experiment were estimated with an automated bandwidth selection method in \cite{Botev2010}. For each retrieval, we used a different $\kappa$, which was selected based on the strategy presented in Section \ref{sc:selectkappa} (by assuming that no training data are available).   From this result, we can see that, the proposed method outperforms the compared similarity measures. Since this classification is only based on $6$ features, with more features, one can potentially improve the classification result
by adding more features.

\begin{table}
\begin{center}
\caption{Classification result of texture images.}
\begin{tabular}{|c|c|c|c|c|c|c|}
\hline
Categories & Hist. ($20$) & Hist. ($100$) & $\ltwo$  & $\chi^2$& Bhatt. & $d_\kappa$\\
\hline
Leaves(9) & 3 & 3 & 5& 5 & 4& 9 \\
\hline
Food(9)& 6 & 6&6 &7 & 6&  7\\
\hline
Fabric(9)& 2 & 2 & 3&3  & 3 & 3\\
\hline
Buildings(9) & 5 & 6 & 6&7  & 7 &  7\\
\hline
Brick(9) & 3 & 3 & 5 &5  & 4&  5\\
\hline
Bark(9) & 4 & 4 & 5 &5  & 3 &  5\\
\hline
Total (\%) & 23 (42.6) & 24 (44.4) & 30 (55.6)&32 (59.3) & 27 (50.0) &  36 (66.7)\\
\hline
\end{tabular}
\end{center}
\label{tab:textureclass}
\end{table}


\subsection{Simulation Studies on $ \s^2$}
\begin{figure}
\begin{center}
\setlength\tabcolsep{1pt}
\begin{tabular}{cc}
\includegraphics[height=2.5 in]{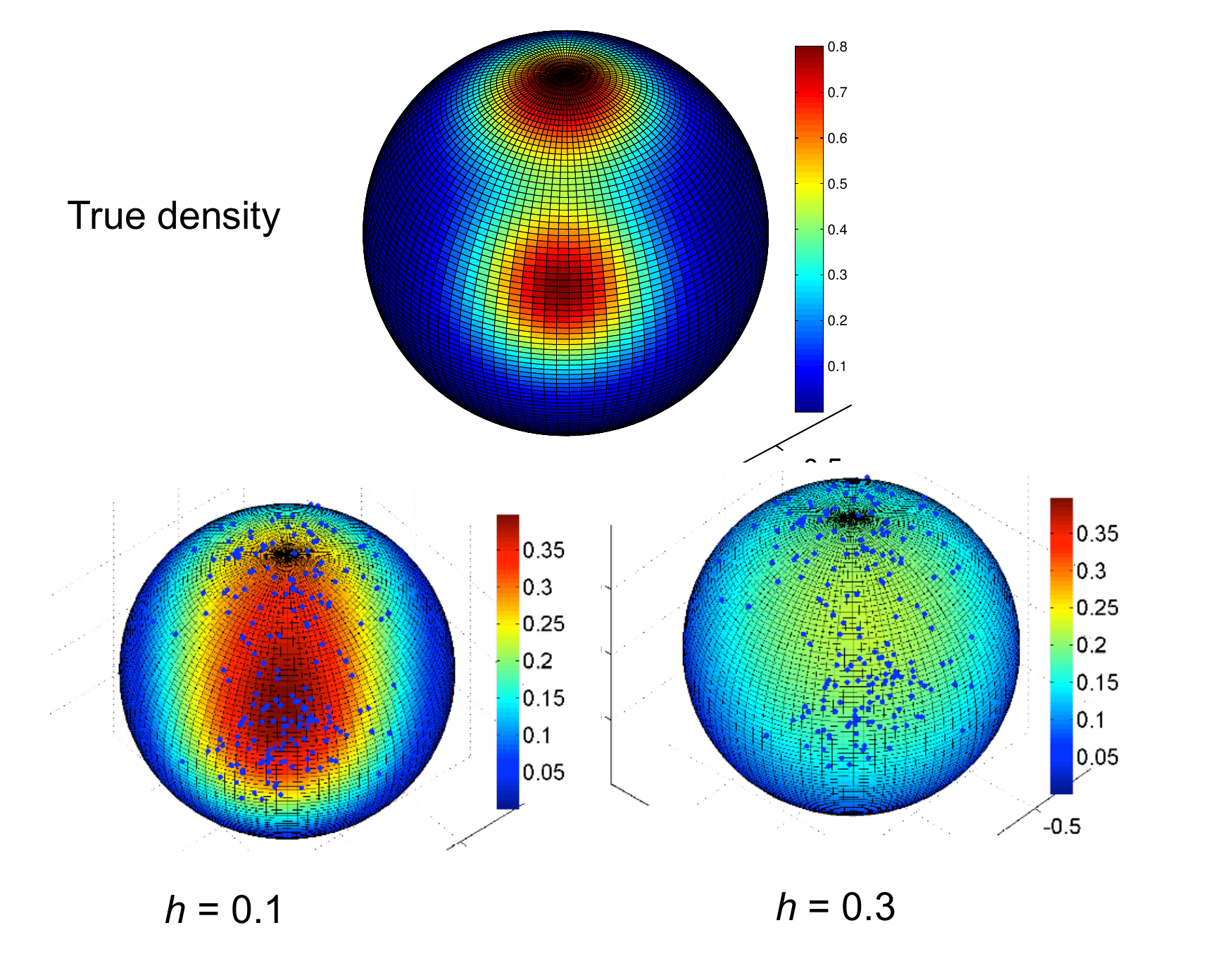}&
\includegraphics[height = 1.5 in]{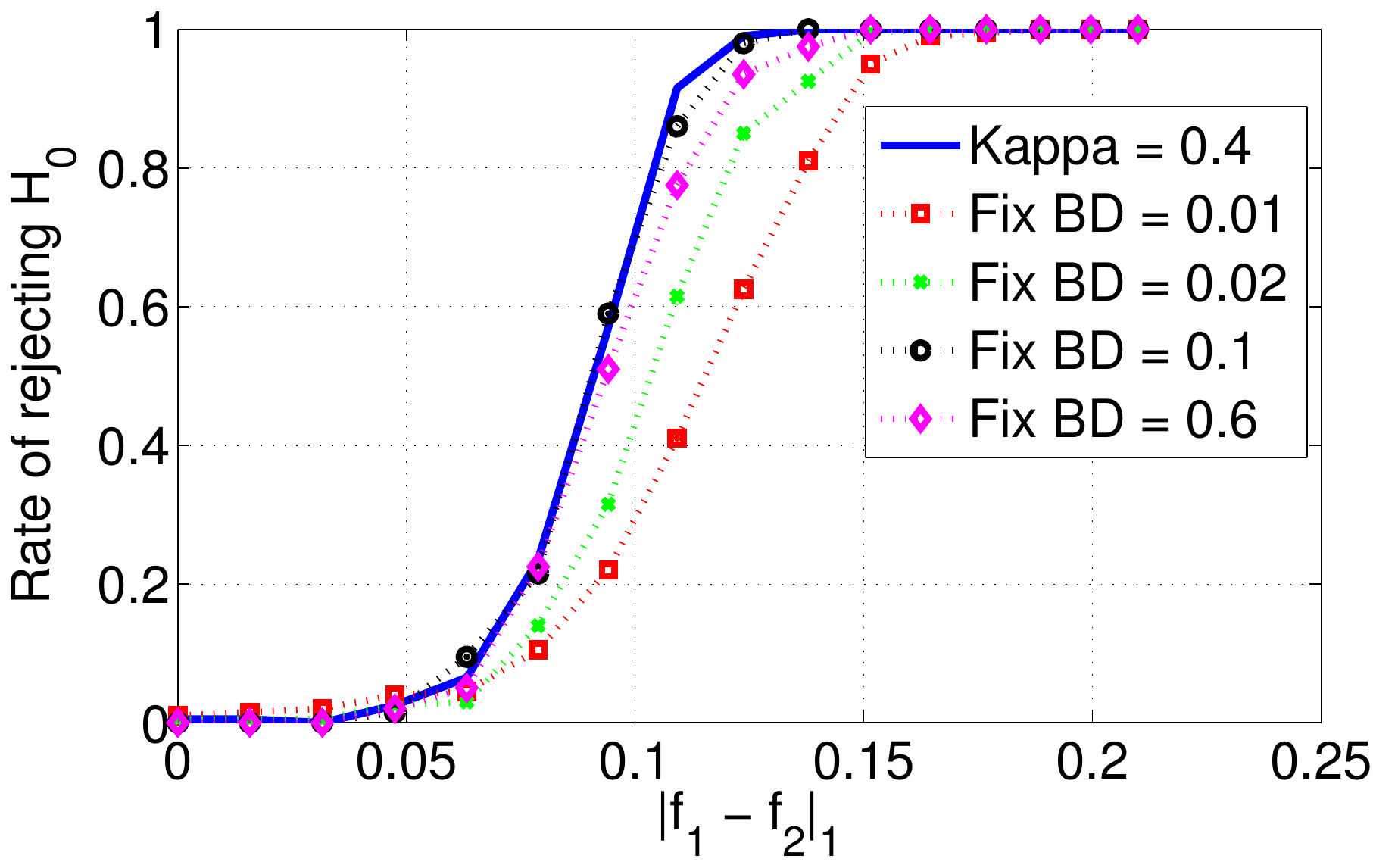}\\
\end{tabular}
\caption{Left: Density estimation on $\mathbb{S}^2$ under different bandwidths. Right: Two-sample hypothesis test on domain $\mathbb{S}^2$.}
\label{fig:densitys2}
\end{center}
\end{figure}

Now we consider the unit two-sphere as the domain of interest. 
We first compare the kernel densities estimated from different random samples. We drew two sets of samples from two different mixtures of Von Mises-Fisher distributions, with sample size of $200$ for each sets.  The heat kernel on $\s^2$ in Eqn. (\ref{eq:sphereke}) was used to estimate  densities.  Note 
that the summation in Eqn. (\ref{eq:sphereke}) has to be truncated in practice; 
only the first $M$ (a large integer) terms were kept to get an approximate Gaussian kernel. In the simulation process, a sphere was parametrized using a $100\times 100$ grid, and a kernel estimated density was fitted using $36$ spherical harmonics basis (up to the degree of $5$). Since the data were simulated from smooth distributions, those spherical harmonic functions are enough to represent the estimated density. If we have a rougher function, more basis elements become necessary. In Figure \ref{fig:densitys2} left panel shows a true density function that we used to sample data, and the estimates with the bandwidths $h=0.1$ and $h=0.3$. Next, we compare $d_\kappa$ with the Fisher-Rao metric, which is defined as $d_{fr}(g_1,g_2) = \cos^{-1} \left( \int_D \sqrt{g_1} \sqrt{g_2} ds \right)$ for any densities $g_1,g_2$ on $\mathbb{S}^2$. The experiment results are shown in Table \ref{tab:diffPD}. Here we used $\kappa = 0.2$ for our approach.  This experiment shows that the proposed distance is almost constant and not affected by the selected bandwidth, in contrast to the Fisher-Rao distance that changes significantly with the bandwidth used.

\begin{table*} \centering
\caption{Comparison of $d_\kappa$ with Fisher-Rao distance. }

\ra{1.3}
\begin{tabular}{@ {}rcrrrrcrrrr@{}} \toprule
& & \multicolumn{4}{c}{$d_\kappa$} &  \phantom{abc} & \multicolumn{4}{c}{Fisher-Rao ($d_{fr}$)} \\
\cmidrule{3-6}  \cmidrule{8-11}
 && \multicolumn{4}{c}{Bandwidth ($f_2$)} &\phantom{abc} & \multicolumn{4}{c}{Bandwidth ($f_2$)}\\ 
 Bandwidth ($f_1$) && 0.05 & 0.1 & 0.15 & 0.2 && 0.05 & 0.1 & 0.15 & 0.2 \\ \midrule
0.05 &&  0.1470  &  0.1478  &  0.1485   & 0.1492 &&  0.4485  &  0.3857  &  0.3937 &   0.4275 \\
0.1 && 0.1453  &  0.1461  &  0.1467   & 0.1475 && 0.4727  &  0.3466  &  0.2998  &  0.3015 \\
0.15 && 0.1440  &  0.1447 &   0.1453  &  0.1459 && 0.5218  &  0.3699  &  0.2847  &  0.2489 \\
0.2 && 0.1423  &  0.1429   & 0.1435 &   0.1439 && 0.5695  &  0.4089  &  0.3045  &  0.2413 \\
\bottomrule
\end{tabular}
\label{tab:diffPD}
\end{table*}

Next, we used $d_\kappa$ to perform a two-sample hypothesis testing on $\mathbb{S}^2$. We simulated $16$ pairs of density functions on $\s^2$ with increasing $\mathbb{L}^1$ distances and sampled $500$ data points from each of them. The bandwidth for density estimation can be selected using a data driven method
given in \cite{klemela2000estimation}.  
The test results are shown in Figure \ref{fig:densitys2} right panel. We have selected a constant $\kappa = 0.4$ in the experiment (similar to the $\s^1$ case).  We only compared with the $\ltwo$ metric since other testing methods (e.g., KS and MMD) are not directly applicable for data on $\s^2$.

\subsection{Real Data Application on $\s^2$}

An interesting application of our approach on $\D = \s^2$  is in analyzing hurricane data for studying patterns of hurricanes. In the result reported here, we used the
Atlantic hurricane database (HURDAT2) \citep{HURDAT2}, which contains hurricanes starting from north Atlantic ocean and Gulf of Mexico. The database contains six-hourly information on the location, maximum winds, central pressure and so on, for each of the relevant hurricanes. 

\begin{figure}
\begin{center}
\begin{tabular}{ccc}
\includegraphics[height = 1.2 in]{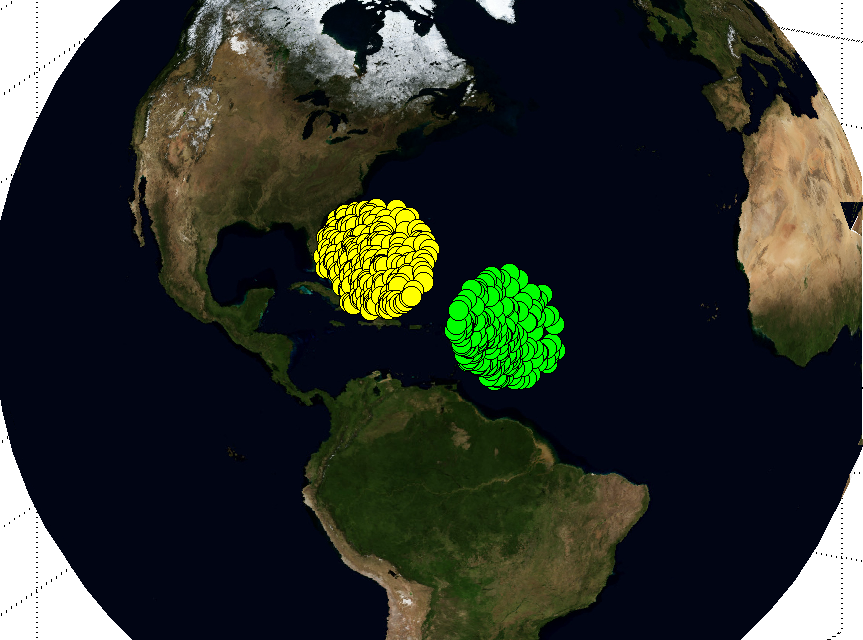}&
\includegraphics[height = 1.2 in]{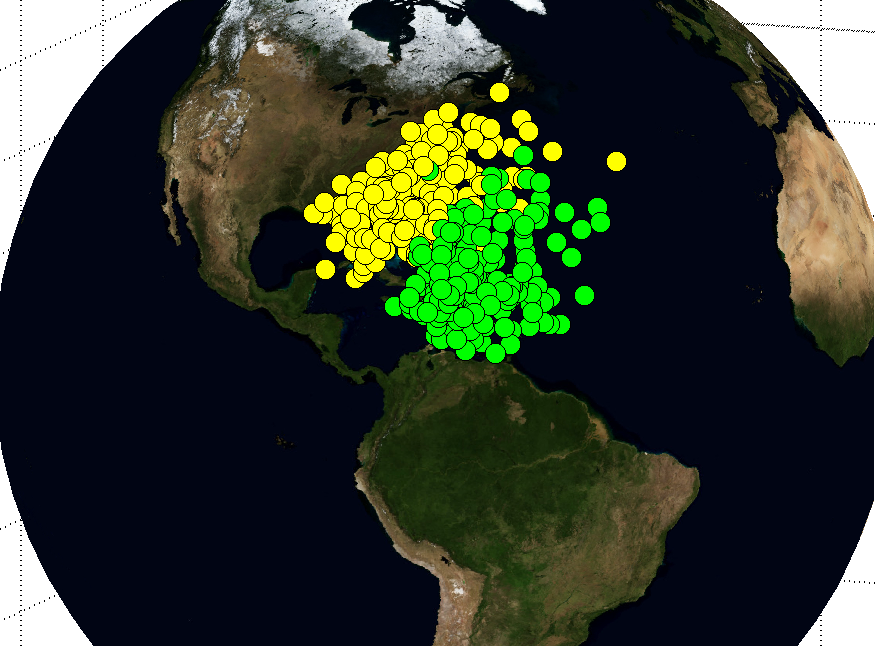}&
\includegraphics[height = 1.2 in]{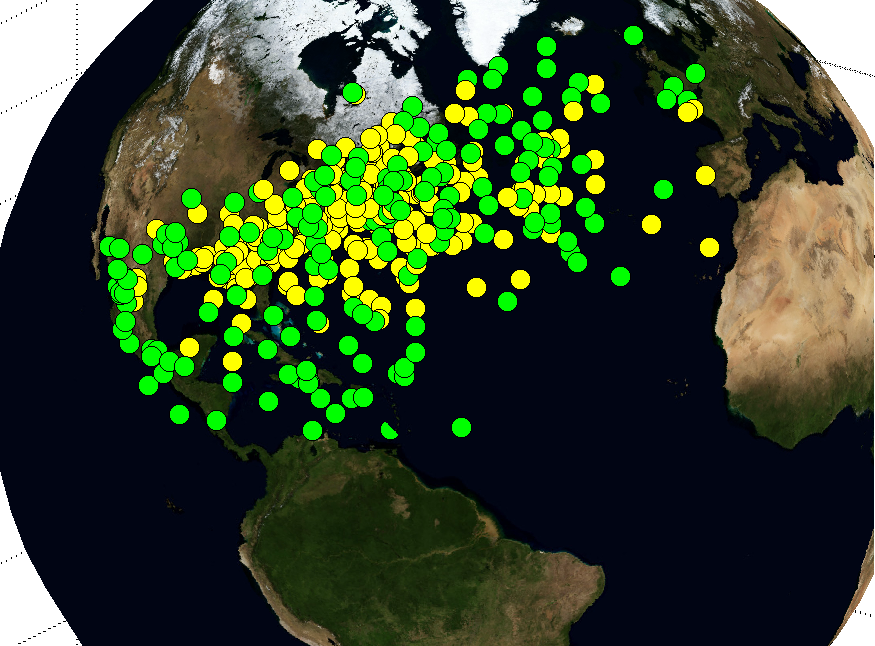}\\
(a) Starting points & (b) After $60$ hours & (c) Ending points\\
\end{tabular}
\caption{Location distribution of hurricanes starting from different regions. }
\label{fig:hurricanetracks}
\end{center}
\end{figure}

First, we are interested in analyzing the location distributions of hurricanes starting from two different regions. In Figure \ref{fig:hurricanetracks}, panel (a) shows two sets of hurricanes according to their starting locations (in different colors), panel (b) shows locations of these hurricanes after $60$ hours, and panel (c) shows the ending points of them. Using $d_\kappa$, we can measure the difference between location distributions of these two sets of hurricanes after a certain period of development.
We can also perform a two-sample hypothesis testing to see if the hurricane location distributions are different after a certain period development. We randomly chose three pairs of sets of hurricanes and calculated $d_\kappa$ for each pair. Table \ref{tab:hurrican} shows the experiment result, where  ``1''  represents rejecting the null hypothesis and  ``0'' represents failing to reject the null hypothesis (based on the significance level $\alpha = 0.05$ ).  All $d_{\kappa}$ and two-sample hypothesis tests were calculated on the section $S_{\kappa = 1}$. From the table we can see that the short-term evolution of hurricanes depends on their starting points; however, as the time lag increases the dependence naturally decreases, and eventually does not depend on the initial locations (e.g., the first and second pair). However, when the initial locations are significantly different, the ending points also are discriminative (e.g., the third pair).  The p-values of two-sample hypothesis tests for the three pairs at the ending stage are $p = 0.42$, $p=0.07$ and $p=0.00$, respectively. 

\begin{figure}
\begin{center}
\begin{tabular}{ccc}
\includegraphics[height = 1.2 in]{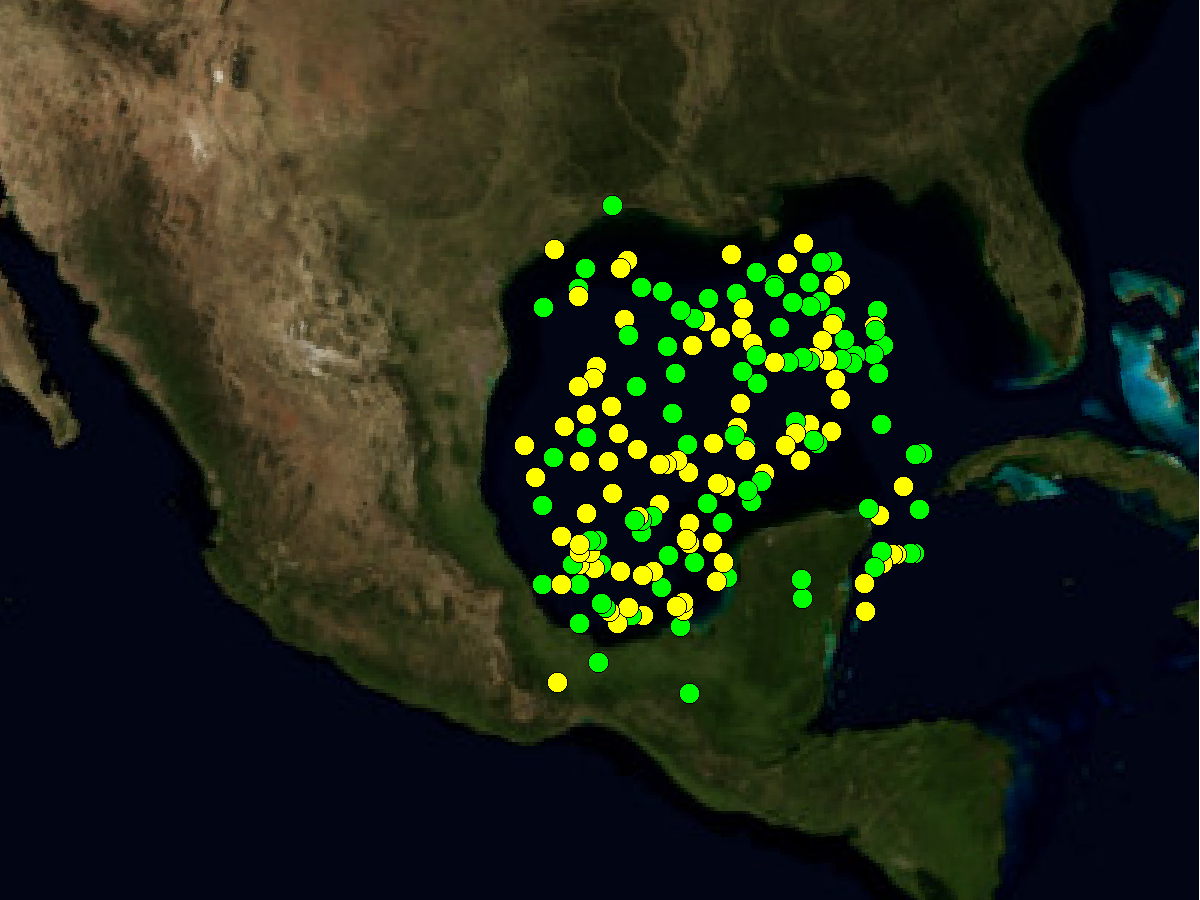}&
\includegraphics[height = 1.2 in]{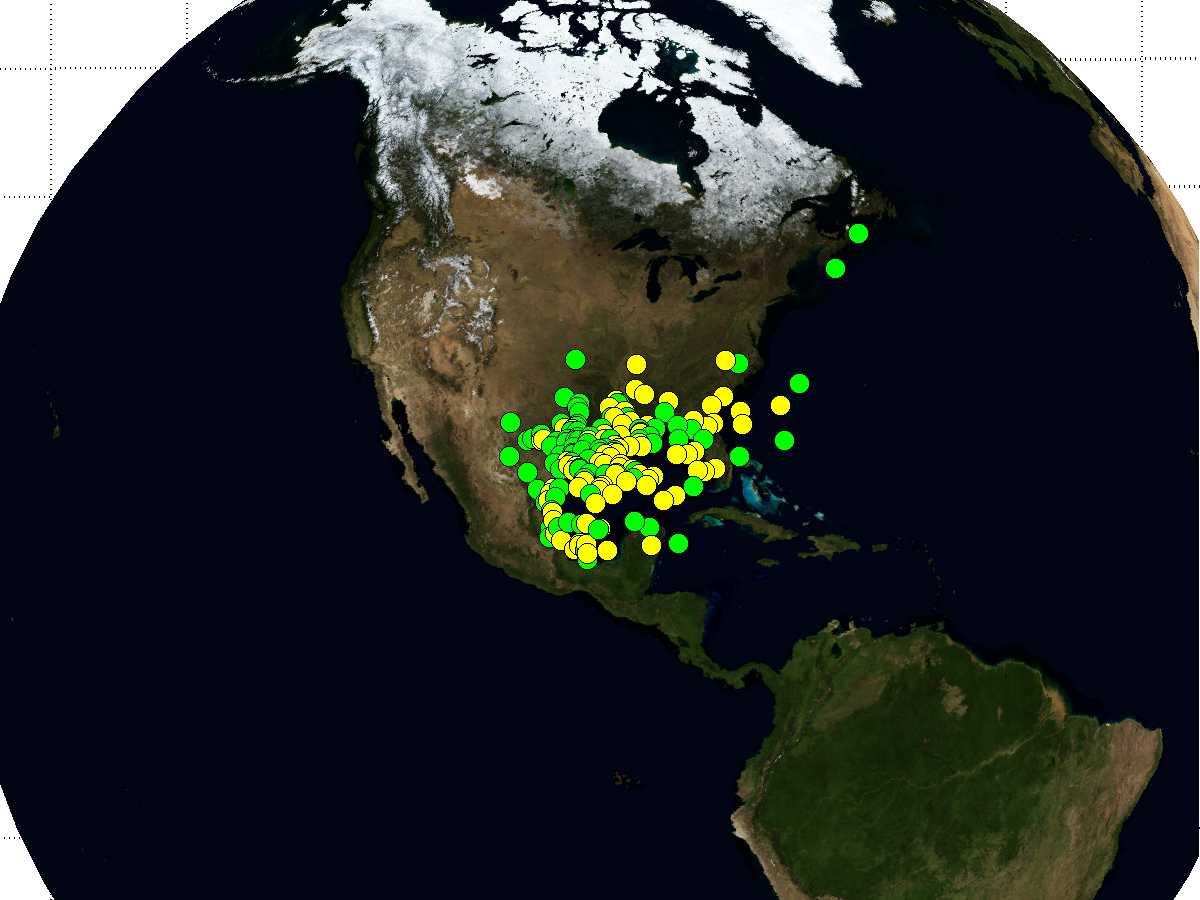}&
\includegraphics[height = 1.2 in]{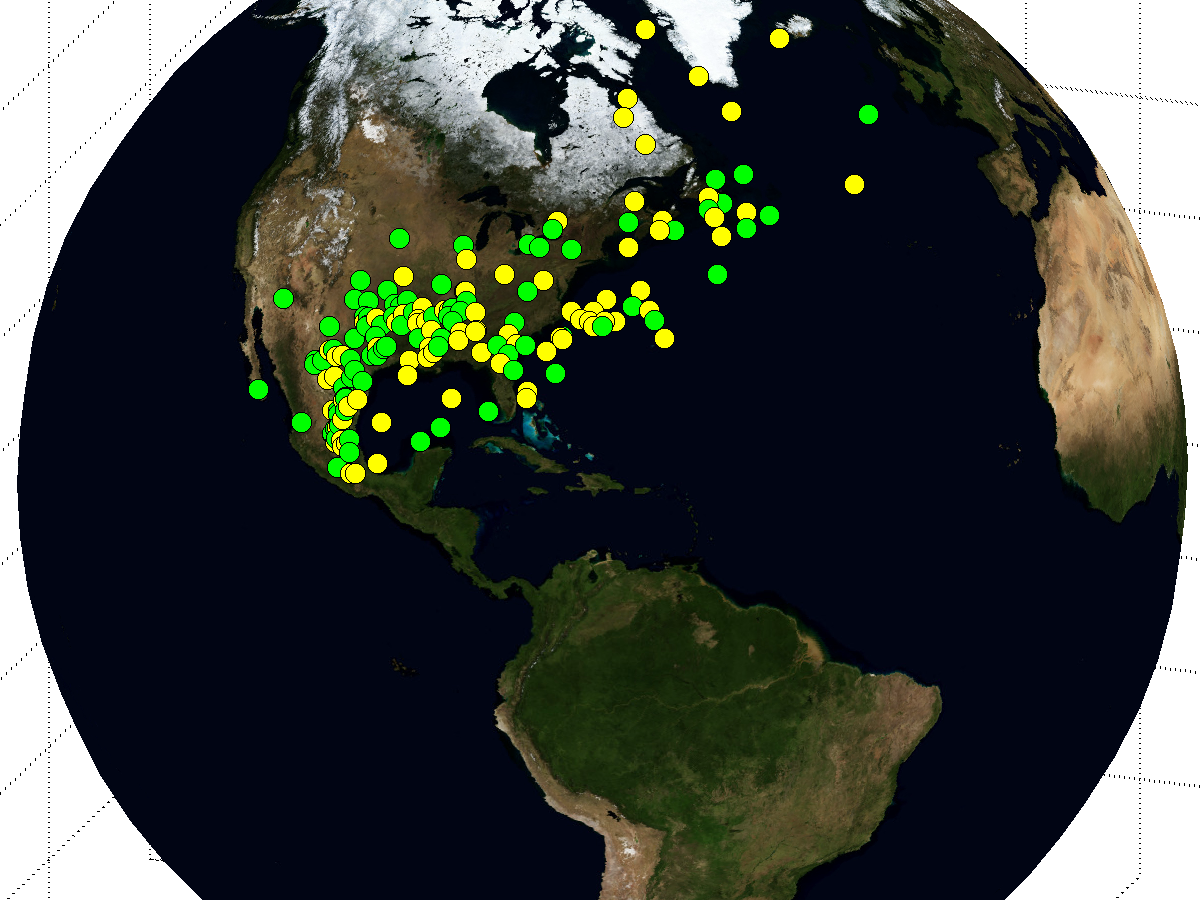}\\
(a) Starting points & (b) After $60$ hours & (c) Ending points\\
\end{tabular}
\caption{Distribution of two sets of hurricanes in Gulf of Mexico. Hurricanes between  [May, August] are marked in yellow and hurricanes between [September, December] are marked in green. }
\label{fig:twoperiod}
\end{center}
\end{figure}

Next, we divided hurricanes starting from the Gulf of Mexico into two categories: (1) hurricanes started in [May, August], and (2) hurricanes started in [September, December], and analyzed their ending points. Figure \ref{fig:twoperiod} shows these two categories of hurricanes in yellow and green color, respectively. In Figure \ref{fig:twoperiod}, panel (a) shows the starting points of these hurricanes, panel (b) shows the hurricane locations after $60$ hours and panel (c) shows the ending points of these hurricanes. Our results indicate that the distributions of starting points of these two sets of hurricanes have no statistical difference.  But after $60$ hours' development, the distributions of them are significantly different, and the distributions of their ending points are also different. From Figure \ref{fig:twoperiod}, we can see that most hurricanes proceed along the east coast of America, and hurricanes in  [May, August] in generate spread out faster and farther than hurricanes in [September, December]. 

\begin{table}
\begin{center}
\caption{Comparison of hurricanes starting at different locations on section $S_{\kappa = 1}$ (1 - reject the null hypothesis; 0 - fail to reject the null hypothesis). }
\begin{tabular}{c|c|c c c c c c c c}
\hline
\multicolumn{2}{c|}{Temporal info} &Starting & 6-hour & 12-hour & 18-hour &  24-hour &  30-hour & 60-hour & Ending  \\
\hline
\multirow{2}{*}{First pair}& $d_{\kappa}$ & 0.1574 & 0.1527 & 0.1476 & 0.1423 & 0.1370 & 0.1318 & 0.1111 & 0.0341 \\
\cline{2-10}
&Test & 1 & 1 & 1 & 1 & 1 & 1 & 1 & 0 \\
\hline
\multirow{2}{*}{Second pair}& $d_{\kappa}$ & 0.2331 & 0.2253 & 0.2176 & 0.2095 & 0.2020 & 0.1948 & 0.1682 & 0.0416 \\
\cline{2-10}
&Test & 1 & 1 & 1 & 1 & 1 & 1 & 1 & 0 \\
\hline
\multirow{2}{*}{Third pair}& $d_{\kappa}$ & 0.4512 & 0.4403 & 0.4277 & 0.4169 & 0.4035 & 0.3915 & 0.3419 & 0.1417 \\
\cline{2-10}
&Test & 1 & 1 & 1 & 1 & 1 & 1 & 1 & 1 \\
\hline
\end{tabular}
\end{center}
\label{tab:hurrican}
\end{table}

 \section{Summary}
We have introduced a framework for metric-based comparison of densities that have 
been estimated using an isotropic Gaussian kernel. This comparison is based on quantifying the smoothness levels of density functions
and bringing them to the same level before performing comparisons. The quantification and manipulation of the smoothing levels of 
{\it pdf}s are built on an action of a smoothing group on the space of functions. This action is implemented with the 
help of the heat equation whose solutions correspond to a Gaussian isotopic smoothing of an initial function. A section
of this action is a set of all functions that have the same level of smoothness and 
this set can be identified with an ellipsoid. Geodesic distances on this ellipsoid provide a measure for 
comparing estimated densities. We use
this framework to derive a two-sample hypothesis test using geodesic distance as a test statistic and 
bootstrap method for approximating distribution for this test statistic. Through a variety of experiments and studies involving 
both real and simulated data, we test the validity of this approach on several domains including a  unit circle, a unit interval, and 
two-dimensional unit sphere. It is observed that the task of bringing estimated densities to the same smoothness
level reduces the effect of bandwidth and/or sample size on density comparisons and significantly improves the
test results. 

 \section{Appendix}
 
 \subsection{Proof that $S_\kappa$ is an Orthogonal Section  } \label{proof:thm1}

An orthogonal section $S_\kappa$ is a subset of $\real^{N+1}$ (coefficient representation of densities) under the action of the  group $\real$ (defined in Eqn. (3) in the main paper)
if: (i) one and only one element of every orbit $[\bc]$ in $\real^{N+1}$ presents in $S_\kappa$, and 
(ii) the set $S_\kappa$ is perpendicular to every orbit at the point of intersection. The last property means
that if $S_\kappa$ intersects an orbit $[\bc]$ at $\tilde{\bc}$, then $T_{\tilde{\bc}}(S_\kappa) \perp T_{\tilde{\bc}}([\bc])$.  We need to verify the two properties: (1) The function $t \mapsto \sum_{n} e^{-2\lambda_n t} \lambda_n c_n^2$ is a strictly 
monotonically-decreasing function that ranges  ($+\infty$, $0$). Thus, for any $\bc \in \real^{N+1}$ 
and $\kappa > 0$, 
there exists a unique $t^*$ such that $\sum_{n} e^{-2\lambda_n t^*} \lambda_n c_n^2 =  \kappa$. 
(2) At any point $\bc \in S_{\kappa}$, the space 
normal to $S_{\kappa}$ (inside $\real^{N}$, notice that $\lambda_0 = 0$)  is a one-dimensional 
space spanned by 
the vector ${\bf n}_{\bc}  = \{ \lambda_1 c_1, \lambda_2 c_2, \dots, \lambda_N c_N\}$. 
Let ${\bf u}_{\bc}$ denote the unit vector in the normal direction 
${\bf u}_{\bc} = {\bf n}_{\bc}/\| {\bf n}_{\bc}\|$. Since $S_{\kappa}$ is 
a level set of $G$, it is automatically perpendicular to ${\bf u}_{\bc}$ and $T_c([\bc])$. In other words, the orbits are just the flow lines for the  
gradient vector field of the function $G$ and since the level sets of a functional 
are perpendicular to the flow lines of gradient of that function, 
it follows that the $S_{\kappa}$ 
is perpendicular to these orbits.
 
  \subsection{Path Straightening Algorithm on $S_\kappa$ } \label{proof:thm1}
Here we present the path straightening algorithm for calculating distances on $S_\kappa$.  We first list the following basic tools for the path straightening algorithm. 

\begin{enumerate}
\item {\bf Projection onto Mainfold $S_{\kappa}$}: 
For any arbitrary point $\bc \in \real^{N}$, we need a tool to project $\bc$ to the nearest point in $S_{\kappa}$. 
One can find this nearest point 
by iteratively updating $\bc$ according to $\bc \mapsto  \bc + (\kappa -  G(\bc)){\bf u}_{\bc}$, 
until $G(\bc) = \kappa$. 

\item {\bf Projection onto the Tangent Space $T_c(S_k)$}:
 Given a vector $w \in \real^{N} $, we need to project  $w$ onto $T_{\bc}(S_{\kappa})$. 
 Since the unit normal to $S_{\kappa}$ at $\bc$ is ${\bf u}_{\bc}$, 
 the projection of $w$ on $T_{\bc}(S_{\kappa})$ is given by 
 $w \rightarrow  ( w - \left\langle w , {\bf u}_{\bc} \right\rangle 
 {\bf u}_{\bc} )$.

\item {\bf Covariant Derivative and Integral:} Let $\alpha$ be a given path on $S_{\kappa}$, i.e., $\alpha:[0,1] \to S_{\kappa}$, and let $w$ 
be a vector field along $\alpha$, i.e., for each $\tau \in [0,1]$, $w(\tau) \in T_{\alpha(\tau)}(S_{\kappa})$. We define
the covariant derivative of $w$ along $\alpha$, denoted $\frac{Dw}{d\tau}$, to be the vector field obtained by projecting $\frac{dw}{d\tau}(\tau)
\in \real^N$ onto 
the tangent space $T_{\alpha(\tau)}(S_{\kappa})$.  
Covariant integral is the inverse procedure of covariant derivative. A vector field $u$ is called a covariant integral of $w$ along $\alpha$ 
if the covariant derivative of $u$ is $w$, i.e., $\frac{Du}{d\tau} = w$. Using the previous item on projection, one 
can derive tools for computing covariant derivatives and integrals of any given vector field. 

\item {\bf Parallel Translation:} We will also 
need tools for forward and backward parallel translation of tangent vectors
along a given path $\alpha$ on $S_{\kappa}$. 
A forward parallel translation of a tangent vector $w \in T_{\alpha(0)}(S_{\kappa})$, is a vector field 
along $\alpha$, denoted $\tilde{w}$, 
such that the covariant derivative of $\tilde {w}$ is $0$ for all $\tau \in [0,1]$, 
i.e., $\frac{D \tilde{w}(\tau)}{d\tau} = 0$, and $\tilde{w}(0) = w$. Similarly, backward parallel translation of a tangent vector 
$w \in T_{\alpha(1)}(S_{\kappa})$, satisfies that $\tilde{w}(1) = w$ and $\frac{D \tilde{w}(\tau)}{d\tau} = 0$ for all $\tau \in [0,1]$.
\end{enumerate}

 \noindent 
{\bf Algorithm} (Path Straightening in $S_{\kappa}$): Given two points $p_1$ and $p_2$ in $S_{\kappa}$. Suppose $p_1,p_2 \in \real^N$, and $\tau = 0,1,2,...,k$.
\begin{enumerate}
\item Initilize a path $\alpha$: for all $\tau = 0,1,2,...k$, using a straight line $(\tau/k)p_1+(1-(\tau/k))p_2$ in $\real^{N}$. 
Project  each of these points to their nearest points in $S_{\kappa}$ to obtain $\alpha(\tau/k)$. 
 
\item Compute $\frac{d \alpha}{d \tau}$ along $\alpha$: let $\tau = 1,2,...,k$ and $v(0) = {\bf 0}$. 
Compute $v(\tau/k) = k(\alpha(\tau/k)-\alpha((\tau-1)/k))$ in $\real^{N}$. Project $v(\tau/k)$ into 
$T_{\alpha(\tau/k)}(S_{\kappa})$ to get $\frac{d \alpha}{dt}(\tau/k)$. 

\item Compute covariant integral of $\frac{d \alpha}{d\tau}$, with zero 
initial condition, along $\alpha$ to obtain a vector field $u$ along $\alpha$. 

\item Backward parallel translate $u(1)$ along $\alpha$ to obtain $\tilde{u}$.  

\item Compute gradient vector field of $E$ according to $w(\tau/k)=u(\tau/k)-(\tau/k)(\tilde{u}(\tau/k))$  for all $\tau$.

\item Update path $\tilde{\alpha}(\tau/k) = \alpha(\tau/k)-\epsilon w(\tau/k)$ by selecting a small $\epsilon>0$. 
Then project $\tilde{\alpha}(\tau/k)$ to $S_{\kappa}$ to obtain the updated path $\alpha(\tau/k)$.

\item Return to step 2 unless $\|w\|$ is small enough or max iteration times reached. 
\end{enumerate}

\bibliography{paper}
\bibliographystyle{ECA_jasa}

\end{document}